\documentclass[%
nofootinbib,
superscriptaddress,
 preprintnumbers,
 amsmath,amssymb,ams
 aps,
]{revtex4-2}

\usepackage{ulem}
\usepackage{bbm}
\usepackage{graphicx}
\usepackage{dcolumn}
\usepackage{bm}

\usepackage{multirow}
\usepackage{pgfplots}
\usepackage{tikz}
\usepackage{tikz-feynman}
\usepackage{slashed}
\usepackage{braket}
\usepackage{cancel}
\usetikzlibrary{shapes.misc}
\tikzset{cross/.style={cross out, draw=black, minimum size=2*(#1-\pgflinewidth), inner sep=0pt, outer sep=0pt},
cross/.default={1pt}}
\begin{document}
\preprint{KEK-TH-2885}
\preprint{J-PARC-TH-0343}
\preprint{RIKEN-iTHEMS-Report-26}
\title{Two-Particle Correlation Function in a Finite Volume}
\date{\today}
\author{Wren Yamada}
\email{rento.yamada@riken.jp}
\affiliation{RIKEN iTHEMS, Wako, Saitama 351-0198, Japan}%

\author{Osamu Morimatsu}
\email{osamu@post.kek.jp}
\affiliation{Theory Center, Institute of Particle and Nuclear Studies (IPNS), High Energy Accelerator Research Organization (KEK), 1-1 Oho, Tsukuba, Ibaraki, 205-0801, Japan}
\affiliation{RIKEN Nishina Center, Wako 351-0198, Japan}%

\author{Toru Sato}
\email{tsato@rcnp.osaka-u.ac.jp}
\affiliation{Research Center for Nuclear Physics (RCNP), the University of Osaka, Ibaraki, Osaka 567-0047, Japan}%

\author{Koichi Yazaki}
\email{koichiyzk@yahoo.co.jp}
\affiliation{RIKEN iTHEMS, Wako, Saitama 351-0198, Japan}%

\begin{abstract}
We  investigate the two-particle correlation function in a finite volume.
First, defining an unphysical complex-energy sheet of the correlation function and identifying unstable states as poles on the sheet, we demonstrate that the expansion of the correlation function, which includes poles of unstable states, holds up to exponentially decreasing corrections with the volume size.
Based on this result, we show that the finite-volume correction to the complex energy of the unstable state also decreases exponentially with the volume size, namely that finite-volume corrections are relatively small for the unstable states.
Then, we show that the imaginary-time correlation function in a finite volume approaches that of the infinite volume as $\tau/L^2 \rightarrow 0$ while it is dominated by the lowest energy mode as $\tau/L^2 \rightarrow \infty$, where $\tau$ is the imaginary time, $L$ is the volume size.
In drawing these conclusions, we first present a qualitative, general discussion and then demonstrate quantitative results using simple nonrelativistic models with a contact interaction (EFT) and a separable potential.
\end{abstract}
\maketitle
\section{Introduction}
In a series of papers \cite{Yamada:2020rpd, Yamada:2021cjo, Yamada:2022xam, Yamada:2021azg, Yamada:2023wqp,Yamada:2025lkc} we have proposed and developed the idea of the pole expansion as a model-independent method to analyze experimental results and/or results of lattice QCD simulation, in particular, for unstable states, where \lq\lq unstable states" generally refer to \lq\lq states" such as resonances or virtual states, corresponding to poles on the unphysical sheet of the complex energy.
In a previous paper \cite{Yamada:2025lkc} we have applied the pole expansion to the two-particle imaginary-time correlation function and proposed it as a method to extract information of unstable states such as masses and widths from the two-particle imaginary-time correlation functions obtained by lattice QCD simulations.
By the Mittag-Leffler theorem \cite{arfken2013mathematical,whittaker1928course}, we express the imaginary-time correlation function in the infinite volume as a sum of pole terms in terms of the uniformization variable, which makes the correlation function single-valued \cite{Newton:book,NEWTON195829,Kato:1965iee}.

A natural question is whether finite-volume effects are significant in the two-particle correlation function because actual lattice QCD simulations are performed with a finite volume.
\par
In Ref.\,\cite{DeWitt:1956be} DeWitt studied the transition from discrete to continuous spectra.
It was shown that the level shift of the spectrum in a finite volume due to the interaction of two particles is proportional to the corresponding scattering phase shift.
It was also argued that the finite-to-infinite volume limit and the complex-to-real energy limit are not interchangeable and the correct infinite-volume limit is to take the finite-to-infinite volume limit first, which means that the infinite-volume limit does not exist on the real-energy axis.
L\"{u}scher studied volume dependence of the energy spectrum in massive quantum field theories for stable particle states in Ref.\,\cite{Luscher:1985dn}, for scattering states in Ref.\,\cite{Luscher:1986pf} and two-particle states on a torus and their relation to the scattering matrix (L\"{u}scher equation) in Ref.\,\cite{Luscher:1990ux}.
L\"{u}scher also discussed signatures of unstable particles in a finite volume in Ref.\,\cite{Luscher:1991cf}.
Based on L\"{u}scher equation many works of hadron resonances in the single-channel have been done using lattice QCD simulations\cite{CP-PACS:2007wro,Feng:2010es,CS:2011vqf,Lang:2011mn,Dudek:2012xn,Alexandrou:2017mpi,Andersen:2018mau,Wilson:2015dqa,Dudek:2016cru,Dudek:2014qha,BaryonScatteringBaSc:2023zvt,BaryonScatteringBaSc:2023ori,Moir:2016srx}.
Recently, by extending the L\"{u}scher equation to coupled channels people have started to apply the lattice QCD simulations to hadron resonances in coupled-channels \cite{Dudek:2014qha,Wilson:2015dqa,Dudek:2016cru,BaryonScatteringBaSc:2023zvt,BaryonScatteringBaSc:2023ori,Moir:2016srx}. 
In Ref.\,\cite{Briceno:2017max,Chen:2022asf,Mai:2022eur} scattering processes and resonances from lattice QCD were reviewed where resonances in the L\"{u}scher formalism in the narrow-width approximation are discussed.
\par
In the present paper we ask the following questions.
\begin{itemize}
\item
How do we identify unstable states in a finite volume in general?
\item
How large is the finite volume effect of unstable states, power, $L^{-3}$, or exponential, $\exp(-\#L)$, as the size of the volume, $L$, increases?
\item
How does the finite-volume imaginary-time correlation function behave?
\item
How can one extract information of unstable states in the infinite volume such as the mass and the width from the imaginary-time correlation function in a finite volume?
\end{itemize}
We first present qualitative, general discussion in the next section and then demonstrate quantitative results using simple nonrelativistic models with a contact interaction (EFT) and a separable potential in the section after the next.
For simplicity we restrict ourselves to a nonrelativistic single-channel system of two particles with the same mass, $m$, but the essence of the following results would be unchanged for more complicated systems of relativistic kinematics, coupled channels and/or different masses.
\section{General qualitative discussion}
\subsection{Correlation function}
In the infinite volume the imaginary-time correlation function, $C (\tau)$, is defined by
\begin{align}
  C (\tau) = \bra{0} {\cal O}(\tau)  {\cal O}^{\dagger} (0)  \ket{0}
  = - \int_{-\infty}^{\infty} \frac{dw}{2\pi} e^{-iw\tau} D(iw)
  = - \int_{\varepsilon_{min}}^\infty \frac{d\omega}{2\pi i} e^{-\omega\tau} {\rm Disc} D(\omega),
  \label{C_inf}
\end{align}
where ${\cal O}(\tau)$ is the interpolating operator of the channel of two particles, $D(\omega)$ is the correlation function as a function of the complex energy, $\omega$.
Continuous scattering states above the threshold and bound states, \{$B$\}, below the threshold, if exist, consist of the complete set of the eigenstates of the system.
Then, $D(\omega)$ is represented (spectral representation) as
\begin{align}
  D (\omega) &= - \sum_{B} \frac{\rho_B}{\omega-\omega_B} - \int_{2m}^\infty  d\omega' \frac{\rho(\omega')}{\omega-\omega'},
  \label{D_inf_sr}
\end{align}
and $C(\tau)$ as
\begin{align}
  C(\tau) &= \sum_{B} \rho_B e^{-\omega_B \tau} + \int_{2m}^\infty d\omega' \rho(\omega') e^{-\omega' \tau}.
  \label{C_inf_sr}
\end{align}
\par
Similarly, in a finite volume the imaginary-time correlation function, $C (\tau,L)$, is defined by 
\begin{align}
  C (\tau,L) = \bra{0} {\cal O}(\tau)  {\cal O}^{\dagger} (0)  \ket{0}_L
  = - \int_{-\infty}^{\infty} \frac{dw}{2\pi} e^{-iw\tau} D(iw,L)
  = - \int_{\varepsilon_{min}}^\infty \frac{d\omega}{2\pi i} e^{-\omega\tau} {\rm Disc} D(\omega,L).
  \label{C_fin}
\end{align}
The continuous energies of the scattering states in the infinite volume become discrete and the spectral representation of $D(\omega,L)$ is given as
\begin{align}
  D(\omega,L) &= - \sum_{n} \frac{\rho(n,L)}{\omega-\omega(n,L)},
  \label{D_fin_sr}
\end{align}
and $C(\tau,L)$ as
\begin{align}
  C(\tau,L) &= \sum_{n} \rho(n,L) e^{-\omega(n,L) \tau}.
  \label{C_fin_sr}
\end{align}
\subsection{Complex-energy Riemann sheet}
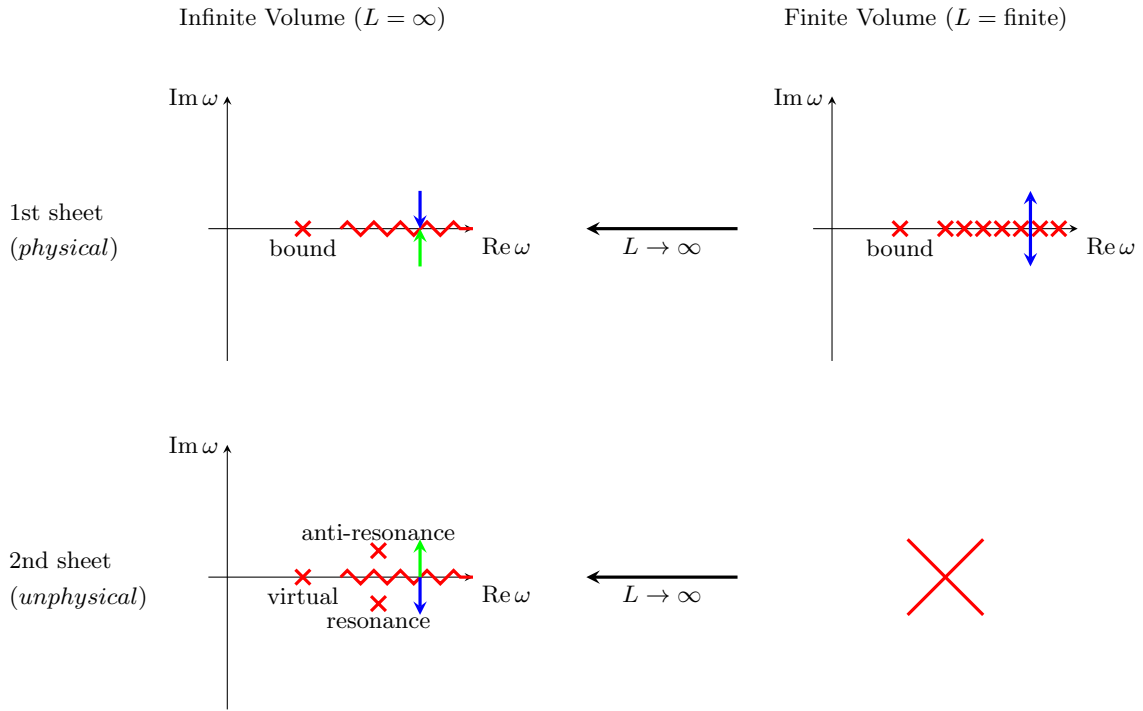
\begin{figure}[h]
	\begin{tikzpicture}[>=stealth]
	\node [above right] at (-3.25,2.5) {Infinite Volume ($L=\infty$)};
	\node [above right] at (-5.5,0) {1st sheet};
	\node [below right]at (-5.5,0) {($physical$)};
	\draw [->] (-2.75,0)--(0.75,0) node[below right] {${\rm Re}\,\omega$};
	\draw [->] (-2.5,-1.75)--(-2.5,1.75) node[left] {${\rm Im}\,\omega$};
	\draw [very thick] (-1.5,0) node [cross=4pt,red]{};
	\node at (-1.5,-0.25) {bound};
	\draw [very thick,red] (-1,0) edge[decorate, decoration=zigzag] (0.75,0);
	\draw [very thick,blue,->] (0.05,0.5)--(0.05,0);
	\draw [very thick,green,->] (0.05,-0.5)--(0.05,0);
	\draw [very thick,<-] (2.25,0)--(4.25,0);
	\node at (3.25,0) [below] {$L \to \infty$};
	\node at (4.75,2.5) [above right] {Finite Volume ($L=\text{finite}$)};
	\draw [->] (5.25,0)--(8.75,0) node[below right] {${\rm Re}\,\omega$};
	\draw [->] (5.5,-1.75)--(5.5,1.75) node[left] {${\rm Im}\,\omega$};	
	\draw [very thick] (6.4,0) node [cross=4pt,red]{};
	\node at (6.4,-0.25) {bound};
	\draw [very thick] (7,0) node [cross=4pt,red]{};
	\draw [very thick] (7.25,0) node [cross=4pt,red]{};
	\draw [very thick] (7.5,0) node [cross=4pt,red]{};
	\draw [very thick] (7.75,0) node [cross=4pt,red]{};
	\draw [very thick] (8,0) node [cross=4pt,red]{};
	\draw [very thick] (8.25,0) node [cross=4pt,red]{};
	\draw [very thick] (8.5,0) node [cross=4pt,red]{};
	\draw [very thick,blue,<->] (8.125,0.5)--(8.125,-0.5);
	\node at (0,-2.5) {};
	\end{tikzpicture}
	\begin{tikzpicture}[>=stealth]
	\node [above right] at (-5.5,0) {2nd sheet};
	\node [below right]at (-5.5,0) {($unphysical$)};
	\draw [->] (-2.75,0)--(0.75,0) node[below right] {${\rm Re}\,\omega$};
	\draw [->] (-2.5,-1.75)--(-2.5,1.75) node[left] {${\rm Im}\,\omega$};
	\draw [very thick,red] (-1,0) edge[decorate, decoration=zigzag] (0.75,0);
	\draw [very thick] (-1.5,0) node [cross=4pt,red]{};
	\node at (-1.5,-0.25) {virtual};
	\draw [very thick] (-0.5,0.35) node [cross=4pt,red]{};
	\node at (-0.5,0.6) {anti-resonance};
	\draw [very thick] (-0.5,-0.35) node [cross=4pt,red]{};
	\node at (-0.5,-0.6) {resonance};
	\draw [very thick,blue,->] (0.05,0)--(0.05,-0.5);
	\draw [very thick,green,->] (0.05,0)--(0.05,0.5);
	\draw [very thick,<-] (2.25,0)--(4.25,0);
	\node at (3.25,0) [below] {$L \to \infty$};
	\phantom{
	\draw [->] (5.25,0)--(8.75,0) node[below right] {${\rm Re}\,\omega$};
	\draw [->] (5.5,-1.75)--(5.5,1.75) node[left] {${\rm Im}\,\omega$};	
	}
	\draw [very thick,red] (6.5,0.5)--(7.5,-0.5);
	\draw [very thick,red] (6.5,-0.5)--(7.5,0.5);
	\end{tikzpicture}
	\caption{Conventional picture of the complex-energy Riemann sheet for the two-particle correlation function in the infinite volume (left) and a finite volume (right).}
	\label{omega_conv}
\end{figure}

In the infinite volume, the Riemann surface of the correlation function consists of two sheets of the complex energy.
Continuous scattering states above the threshold correspond to a branch cut connecting the two sheets (the upper and lower left ones in Fig.\ref{omega_conv}), bound states below the threshold to poles on the physical sheet (the upper left one in Fig.1) and unstable states to poles on the unphysical sheet (the upper left one in Fig.\ref{omega_conv}).
\clearpage
\begin{figure}[h!]
	\begin{tikzpicture}[>=stealth]
	\node [above right] at (-3.25,2.5) {Infinite Volume ($L=\infty$)};
	\node [above right] at (-5.5,0) {1st sheet};
	\node [below right]at (-5.5,0) {($physical$)};
	\draw [->] (-2.75,0)--(0.75,0) node[below right] {${\rm Re}\,\omega$};
	\draw [->] (-2.5,-1.75)--(-2.5,1.75) node[left] {${\rm Im}\,\omega$};
	\draw [very thick] (-1.5,0) node [cross=4pt,red]{};
	\node at (-1.5,-0.25) {bound};
	\draw [very thick,red] (-1,0) edge[decorate, decoration=zigzag] (0.75,0);
	\draw [very thick,blue,->] (0.05,0.5)--(0.05,0);
	\draw [very thick,green,->] (0.05,-0.5)--(0.05,0);
	\draw [very thick,<-] (2.25,0)--(4.25,0);
	\node at (3.25,0) [below] {$L \to \infty$};
	\node at (4.75,2.5) [above right] {Finite Volume ($L=\text{finite}$)};
	\draw [->] (5.25,0)--(8.75,0) node[below right] {${\rm Re}\,\omega$};
	\draw [->] (5.5,-1.75)--(5.5,1.75) node[left] {${\rm Im}\,\omega$};	
	\draw [very thick] (6.4,0) node [cross=4pt,red]{};
	\node at (6.4,-0.25) {bound};
	\draw [very thick] (7,0) node [cross=4pt,red]{};
	\draw [very thick] (7.25,0) node [cross=4pt,red]{};
	\draw [very thick] (7.5,0) node [cross=4pt,red]{};
	\draw [very thick] (7.75,0) node [cross=4pt,red]{};
	\draw [very thick] (8,0) node [cross=4pt,red]{};
	\draw [very thick] (8.25,0) node [cross=4pt,red]{};
	\draw [very thick] (8.5,0) node [cross=4pt,red]{};
	\draw [very thick,blue,<->] (8.125,0.5)--(8.125,-0.5);
	\node at (0,-2.5) {};
	\end{tikzpicture}
	\begin{tikzpicture}[>=stealth]
	\node [above right] at (-5.5,0) {2nd sheet};
	\node [below right]at (-5.5,0) {($unphysical$)};
	\draw [->] (-2.75,0)--(0.75,0) node[below right] {${\rm Re}\,\omega$};
	\draw [->] (-2.5,-1.75)--(-2.5,1.75) node[left] {${\rm Im}\,\omega$};
	\draw [very thick,red] (-1,0) edge[decorate, decoration=zigzag] (0.75,0);
	\draw [very thick] (-1.5,0) node [cross=4pt,red]{};
	\node at (-1.5,-0.25) {virtual};
	\draw [very thick] (-0.5,0.35) node [cross=4pt,red]{};
	\node at (-0.5,0.6) {anti-resonance};
	\draw [very thick] (-0.5,-0.35) node [cross=4pt,red]{};
	\node at (-0.5,-0.6) {resonance};
	\draw [very thick,blue,->] (0.05,0)--(0.05,-0.5);
	\draw [very thick,green,->] (0.05,0)--(0.05,0.5);
	\draw [very thick,<-] (2.25,0)--(4.25,0);
	\node at (3.25,0) [below] {$L \to \infty$};
	\draw [->] (5.25,0)--(8.75,0) node[below right] {${\rm Re}\,\omega$};
	\draw [->] (5.5,-1.75)--(5.5,1.75) node[left] {${\rm Im}\,\omega$};	
	\draw [very thick] (7,0) node [cross=4pt,red]{};
	\draw [very thick] (7.25,0) node [cross=4pt,red]{};
	\draw [very thick] (7.5,0) node [cross=4pt,red]{};
	\draw [very thick] (7.75,0) node [cross=4pt,red]{};
	\draw [very thick] (8,0) node [cross=4pt,red]{};
	\draw [very thick] (8.25,0) node [cross=4pt,red]{};
	\draw [very thick] (8.5,0) node [cross=4pt,red]{};
	\draw [very thick] (6.4,0) node [cross=4pt,red]{};
	\node at (6.4,-0.25) {virtual};
	\draw [very thick] (7.5,0.35) node [cross=4pt,red]{};
	\node at (7.5,0.6) {anti-resonance};
	\draw [very thick] (7.5,-0.35) node [cross=4pt,red]{};
	\node at (7.5,-0.6) {resonance};
	\draw [very thick,green,<->] (8.125,0.5)--(8.125,-0.5);
	\end{tikzpicture}
	\caption{Our picture of the complex-energy Riemann sheet for the two-particle correlation function in the infinite volume (left) and a finite volume (right).}
	\label{omega_new}
\end{figure}
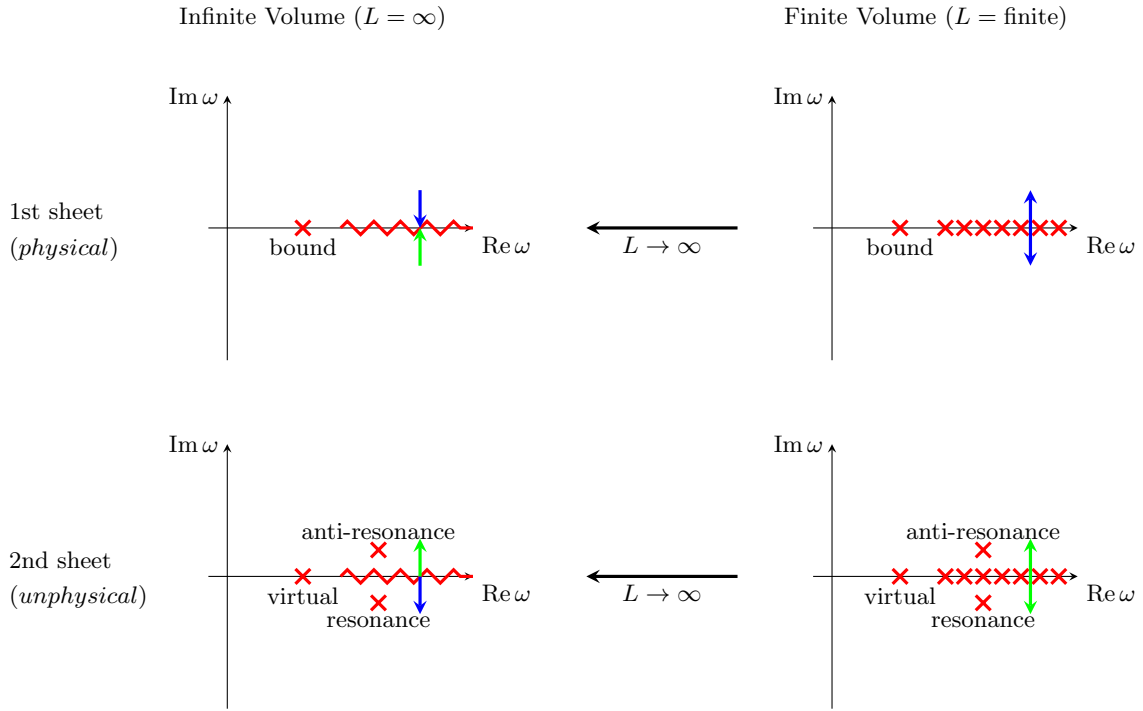

In a finite volume, there is no cut on the complex energy plane because of the absence of the continuous scattering states.
Thus, only one, physical, complex-energy sheet (the upper right one in Fig.\ref{omega_conv}) is conventionally considered for the correlation function.
As will be explicitly demonstrated with using models we obtain the finite-volume correlation function by replacing the integral over the momentum of each particle in the two-particle loop function in the infinite volume with a sum.
The finite-volume correlation function on the physical sheet (the upper right one in Fig.\ref{omega_conv}) converges to the infinite-volume correlation function on the physical sheet (the upper left one in Fig.\ref{omega_conv}) as the volume increases, except exactly on the real axis.

Unlike the conventional approach, we define another, unphysical, complex-energy sheet in a finite volume just as the physical sheet by replacing the integral in the infinite volume by a sum, which will be explicitly demonstrated later by using simple models.
The finite-volume correlation function on the unphysical sheet (the lower right one in Fig.\ref{omega_new}) converges to the infinite-volume correlation function on the unphysical sheet (the lower left one in Fig.\ref{omega_new}) as the volume increases, except exactly on the real axis.
In the infinite volume above the two-particle threshold, the correlation function just above the real axis on the physical sheet is continuous to the function just below the real axis on the unphysical sheet but not the physical sheet.
In contrast, in a finite volume the correlation function just above the real axis on the physical sheet is continuous except for discrete poles to the function just below the real axis on the physical sheet but not the unphysical sheet.
Due to the discontinuity of the physical and unphysical sheets in a finite volume, one might consider that the introduction of the unphysical sheet in a finite volume is meaningless.
We, however, will show that the finite-volume correlation function is expressed as a sum of pole terms on the unphysical sheet together with the bound-state pole terms on the physical sheet, if exist, up to corrections which decrease exponentially as the volume increases.
Therefore, the correlation function on the physical sheet is regarded as related to the function on the unphysical sheet even in a finite volume.
\subsection{Pole expansion}
\begin{figure}[b]
	\begin{tikzpicture}[>=stealth]
	\node at (-2.5,2.5) [red] {$C$};
	\node at (3,3) {$physical$};
	\draw [->] (-3.5,0)--(3.5,0) node[below] {${\rm Re}\,k$};
	\draw [->] (0,-3.5)--(0,3.5) node[left] {${\rm Im}\,k$};
	\draw [very thick] (0,0.5) node [cross=4pt,red]{};
	\node at (0,0.5) [above right] {bound};
	\draw [->,very thick,red] (0,0.7) arc [start angle = 90, end angle = -270, radius = 0.2];
	\node at (3,-3) {$unphysical$};
	\draw[->,very thick,dashed,red] (3,0) arc [start angle = 0, end angle = 360, radius = 3];
	\draw [very thick] (1.5,-0.5) node [cross=4pt,red]{};
	\node at (1.5,-0.6) [below] {resonance};
	\draw [very thick] (0,-0.75) node [cross=4pt,red]{};
	\node at (0,-0.85) [below] {virtual};
	\draw [->,very thick,red] (0,-0.55) arc [start angle = 90, end angle = -270, radius = 0.2];
	\draw [->,very thick,red] (1.5,-0.3) arc [start angle = 90, end angle = -270, radius = 0.2];
	\draw [very thick] (-1.5,-0.5) node [cross=4pt,red]{};
	\node at (-1.5,-0.6) [below] {anti-resonance};
	\draw [->,very thick,red] (-1.5,-0.3) arc [start angle = 90, end angle = -270, radius = 0.2];
	\end{tikzpicture}
\caption{Complex momentum plane for the correlation function in the infinite volume, the upper half plane and the lower half plane correspond to the physical sheet (left) and unphysical sheet (right).}
\label{Contour_inf}
\end{figure}
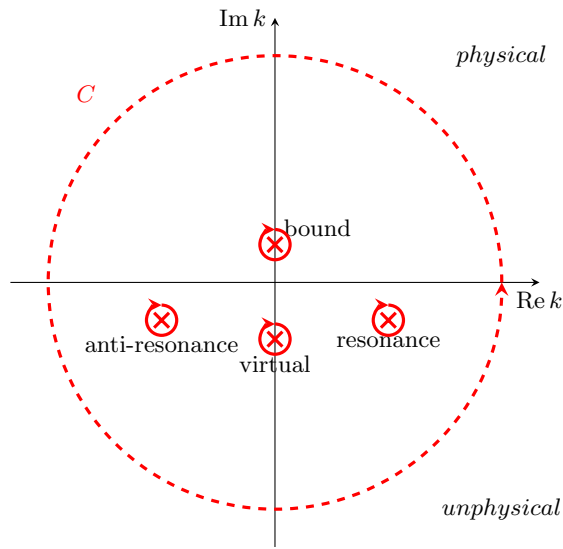
In a nonrelativistic single-channel system of two particles, the momentum, $k$, of each particle in the center-of-mass system plays a role of the uniformization variable, which is related to the energy of the system, $\omega$, as $\omega=2m+k^2/m$.
Then, thanks to Mittag-Leffler theorem, the correlation function, $D(\omega)$, can be expressed in terms of the momentum, $k$, as
\begin{align}
  D(\omega) &= \sum_n \left( \frac{r_n}{k(\omega) - k_n} - \frac{r_n^{*}}{k(\omega) + k_n^{*}} \right),
  \label{D_inf_pole}
\end{align}
up to a subtraction term, which is an entire function and will not be shown explicitly hereafter.
Then,  $C(\tau)$ is expressed as
\begin{align}
  C(\tau) &= \sum_n \left( r_n C (\tau,k_n) - r_n^{*} C (\tau,-k_n^*) \right),
\end{align}
where
\begin{align}
  C(\tau,k_n) = \int_{\varepsilon_{min}}^\infty \frac{d\omega}{2\pi i} e^{-\omega\tau} {\rm Disc} \frac{1}{k(\omega) - k_n}.
\end{align}
We can understand that we obtain Eq.\,\eqref{D_inf_pole} by integrating $\displaystyle{\frac{1}{2\pi i}\frac{D(k')}{k'-k}}$ over $k'$ along the contour $C$ in Fig.\,\ref{Contour_inf}.
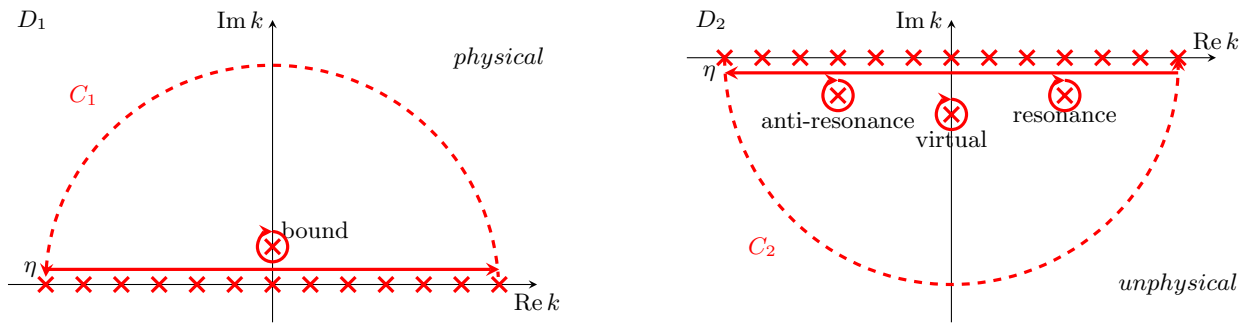
\begin{figure}[b]
	\begin{minipage}{0.495\columnwidth}
	\begin{tikzpicture}[>=stealth]
	\node at (-3.5,3.5) [right] {$D_1$};
	\node at (-2.5,2.5) [red] {$C_1$};
	\node at (3,3) {$physical$};
	\draw [->] (-3.5,0)--(3.5,0) node[below] {${\rm Re}\,k$};
	\draw [->] (0,-0.5)--(0,3.5) node[left] {${\rm Im}\,k$};
	\draw[<-,very thick,dashed,red] (-3,0.1) arc [start angle = 176.186, end angle = 3.81407, radius = 3];
	\draw [->,very thick,red] (-3,0.2)--(3,0.2) ;
	\node at (-3,0.2) [left] {$\eta$};
	\draw [very thick] (0,0.5) node [cross=4pt,red]{};
	\node at (0,0.5) [above right] {bound};
	\draw [->,very thick,red] (0,0.7) arc [start angle = 90, end angle = -270, radius = 0.2];
	\draw [very thick] (-3,0) node [cross=4pt,red]{};
	\draw [very thick] (-2.5,0) node [cross=4pt,red]{};
	\draw [very thick] (-2,0) node [cross=4pt,red]{};
	\draw [very thick] (-1.5,0) node [cross=4pt,red]{};
	\draw [very thick] (-1,0) node [cross=4pt,red]{};
	\draw [very thick] (-0.5,0) node [cross=4pt,red]{};
	\draw [very thick] (0,0) node [cross=4pt,red]{};
	\draw [very thick] (0.5,0) node [cross=4pt,red]{};
	\draw [very thick] (1,0) node [cross=4pt,red]{};
	\draw [very thick] (1.5,0) node [cross=4pt,red]{};
	\draw [very thick] (2,0) node [cross=4pt,red]{};
	\draw [very thick] (2.5,0) node [cross=4pt,red]{};
	\draw [very thick] (3,0) node [cross=4pt,red]{};
	\end{tikzpicture}
	\end{minipage}
	\begin{minipage}{0.495\columnwidth}
	\begin{tikzpicture}[>=stealth]
	\node at (-3.5,0.5) [right] {$D_2$};
	\node at (-2.5,-2.5) [red] {$C_2$};
	\node at (3,-3) {$unphysical$};
	\draw [->] (-3.5,0)--(3.5,0) node[above] {${\rm Re}\,k$};
	\draw [->] (0,-3.5)--(0,0.5) node[left] {${\rm Im}\,k$};
	\draw[->,very thick,dashed,red] (-3,0) arc [start angle = 180, end angle = 360, radius = 3];
	\draw [<-,very thick,red] (-3,-0.2)--(3,-0.2) ;
	\node at (-3,-0.2) [left] {$\eta$};
	\draw [very thick] (1.5,-0.5) node [cross=4pt,red]{};
	\node at (1.5,-0.6) [below] {resonance};
	\draw [very thick] (0,-0.75) node [cross=4pt,red]{};
	\node at (0,-0.85) [below] {virtual};
	\draw [->,very thick,red] (0,-0.55) arc [start angle = 90, end angle = -270, radius = 0.2];
	\draw [->,very thick,red] (1.5,-0.3) arc [start angle = 90, end angle = -270, radius = 0.2];
	\draw [very thick] (-1.5,-0.5) node [cross=4pt,red]{};
	\node at (-1.5,-0.6) [below] {anti-resonance};
	\draw [->,very thick,red] (-1.5,-0.3) arc [start angle = 90, end angle = -270, radius = 0.2];
	\draw [very thick] (-3,0) node [cross=4pt,red]{};
	\draw [very thick] (-2.5,0) node [cross=4pt,red]{};
	\draw [very thick] (-2,0) node [cross=4pt,red]{};
	\draw [very thick] (-1.5,0) node [cross=4pt,red]{};
	\draw [very thick] (-1,0) node [cross=4pt,red]{};
	\draw [very thick] (-0.5,0) node [cross=4pt,red]{};
	\draw [very thick] (0,0) node [cross=4pt,red]{};
	\draw [very thick] (0.5,0) node [cross=4pt,red]{};
	\draw [very thick] (1,0) node [cross=4pt,red]{};
	\draw [very thick] (1.5,0) node [cross=4pt,red]{};
	\draw [very thick] (2,0) node [cross=4pt,red]{};
	\draw [very thick] (2.5,0) node [cross=4pt,red]{};
	\draw [very thick] (3,0) node [cross=4pt,red]{};
	\end{tikzpicture}
	\end{minipage}
\caption{Complex momentum plane for the correlation function in a finite volume, physical sheet (left) and unphysical sheet (right).}
\label{Contour_fin}
\end{figure}
\par
In a finite volume, the spectral representation, Eq.\,\eqref{D_fin_sr}, is also regarded as the pole expansion in the energy, $\omega$, which, however, does not converge to the pole expansion in the infinite volume, Eq.\,\eqref{D_inf_pole}, as the size of the volume increases.

By changing the variable from the complex energy, $\omega$, to the complex momentum, $k$, the physical and unphysical complex-energy sheets are mapped onto the upper and lower half of different complex momentum planes. 
We note the correlation functions on the physical and unphysical complex-energy sheets as $D_1(k,L)$ and $D_2(k,L)$, respectively, and define $D(k,L)$,  by
\begin{align}
  D(k,L) &\equiv
  \begin{cases}
  D_1(k,L) & ({\rm Im} \, k > 0) \\
  D_2(k,L) & ({\rm Im} \, k < 0) 
  \end{cases}
  ,
\end{align}
which approaches the infinite-volume correlation function, $D(k)$, as the volume increases.
By integrating $\displaystyle{\frac{1}{2\pi i}\frac{D(k',L)}{k'-k}}$ over $k'$ along the contour $C_1$ and $C_2$, respectively, on different complex momentum planes in Fig.\,\ref{Contour_fin} and adding the results we obtain
\begin{align}
  D(k,L)
  & = \sum_{n} \left( \frac{r_n(L)}{k-k_n(L)} - \frac{r_n(L)^*}{k+k_n(L)^*} \right) + \frac{1}{2 \pi i} \int_{-\infty}^{\infty}  dk' \frac{D(k'+i\eta,L)-D(k'-i\eta,L)}{k'-k}.
\label{D_fin_pole}
\end{align}
On the right-hand-side of Eq.\,\eqref{D_fin_pole} the integral can be rewritten as
\begin{align}
  \int_{-\infty}^{\infty}  dk' \frac{D(k'+i\eta,L)-D(k'-i\eta,L)}{k'-k}
  & = \int_{-\infty}^{\infty}  dk' \frac{\Delta D(k'+i\eta,L)-\Delta D(k'-i\eta,L)}{k'-k}
\end{align}
where $\Delta D(k,L)$ is the finite-volume correction, $\Delta D(k,L) = D(k,L) - D(k)$.
$\Delta D(k,L)$ is interpreted as due to wrapping-around effects of the particle and the leading behavior of $\Delta D(k,L)$ is proportional to $e^{ikL}$.
One might naively think that the last term of Eq.\,\eqref{D_fin_pole} is $O(e^{-\eta L})$.
However, the path of the integral can be shifted without changing the value of the integral as
\begin{align}
  \int_{-\infty}^{\infty}  dk' \frac{\Delta D(k'+i\eta,L)-\Delta D(k'-i\eta,L)}{k'-k}
  & = \int_{-\infty}^{\infty}  dk' \frac{\Delta D(k'+i\eta',L)-\Delta D(k'-i\eta',L)}{k'-k},
\end{align}
unless the shift of the integral path crosses a pole.
Therefore, the last term of Eq.\,\eqref{D_fin_pole} is $O(e^{- d L})$ where $d$ is the minimum of the absolute value of the imaginary part of $k, k_1, \cdots$ and
\begin{align}
  D(k,L)
  & = \sum_{n} \left( \frac{r_n(L)}{k-k_n(L)} - \frac{r_n(L)^*}{k+k_n(L)^*} \right) + O(e^{-d L}).
\end{align}
This means that in a finite volume the correlation function on the physical sheet is determined by poles on the unphysical sheet as in the infinite volume up to corrections which decays exponentially as the size of the volume, $L$, increases and justifies the idea of the unphysical sheet in a finite volume.

Once the finite-volume correlation function on the unphysical sheet is defined and the unstable states are identified as poles on the unphysical sheet, it is straightforward to obtain the volume dependence of unstable states, which is expected to be exponentially small as the size of the volume increases as the volume dependence of bound states.
\subsection{Two domains of imaginary-time correlation function}
\begin{figure}[b]
	\begin{tikzpicture}[>=stealth]
	\draw [->,very thick] (0,0)--(9,0) node [below] {$\tau$};
	\draw [->,very thick] (0,0)--(0,7) node [left] {$L$};
	\draw [very thick] (0,0)--(8,6);
	\draw [very thick,samples=100,domain=0:8] plot(\x,{3*sqrt(\x/4)});
       \draw (7,3) node [above right] {$\displaystyle{\frac{L^2}{\tau}=\text{const.}}$}; 
	\draw [very thick] (0,0)--(8,6);
       \draw (7,5.25) node [below right] {$\displaystyle{\frac{L}{\tau}=\text{const.}}$}; 
	\draw [->,very thick] (4.5,3)--(7.5,3);
       \draw (6,3) node [below] {$\mu \to \infty$}; 
	\draw [<-,very thick] (0.5,3)--(3.5,3);
       \draw (2,3) node [below] {$\mu \to 0$}; 
	\draw [->,very thick] (4.5,3.875)--(7.5,6.125);
       \draw (5.75,5) node [left] {$\mu \to 0$}; 
	\draw [->,very thick] (4,3.875)--(4,6.125);
       \draw (4,5) node [left] {$\mu \to 0$}; 
	\draw [->,very thick] (4,2.125)--(4,0.5);
       \draw (4,1.3125) node [left] {$\mu \to \infty$}; 
	\end{tikzpicture}
\caption{Asymptotic behavior of the imaginary-time correlation function in the $\tau L$ plane. $\mu \rightarrow 0$ is the infinite volume limit and $\mu \rightarrow \infty$ is the lowest mode dominance.}
\label{C_tauL}
\end{figure}
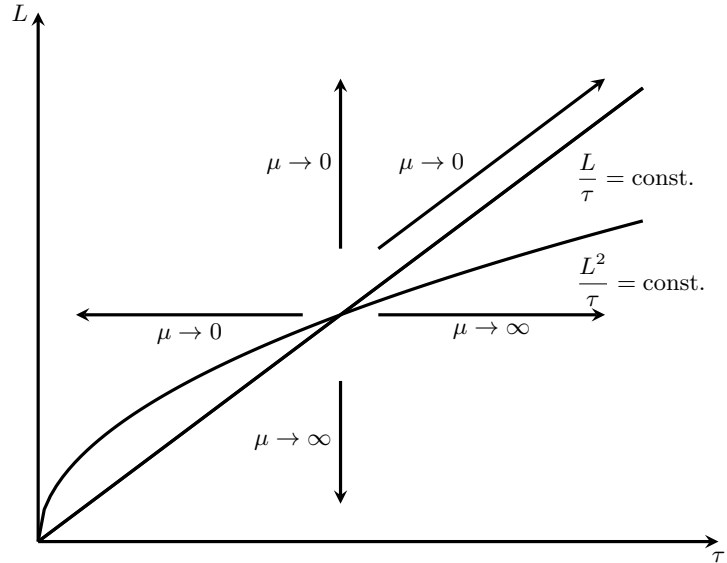
The distance of eigen-energies of the interacting two-particle system is roughly given by that of the noninteracting two-particle system, $\Delta \omega = \frac{4\pi^2}{mL^2}$, and $\Delta \omega \tau = \frac{4\pi^2\tau}{mL^2}=4\pi^2 \mu$, where $\mu$ is a dimensionless parameter $\mu = \frac{\tau}{mL^2}$. The different behaviors of the imaginary-time correlation function is controlled by the value of the dimensionless parameter, $\mu$.
\subsubsection{$\mu \rightarrow \infty$}
As $\mu \rightarrow \infty$, the finite-volume imaginary-time correlation function, $C (\tau,L)$, is dominated by the lowest mode as
\begin{align}
  C (\tau,L) = \rho(0,L) e^{-\omega(0,L) \tau} \left( 1 + O(e^{-4\pi^2 \mu}) \right)
  \rightarrow \rho(0,L) e^{-\omega(0,L) \tau}.
\end{align}
\subsubsection{$\mu \rightarrow 0$}
In order to obtain the asymptotic behavior as $\mu \rightarrow 0$ we consider the finite-volume correction, $\Delta C(\tau,L) = C(\tau,L) - C(\tau)$, which satisfies
\begin{align}
  \Delta C (\tau,L)  = - \int \frac{d\omega}{2\pi i} e^{-\omega\tau} \Delta D(\omega,L).
\end{align}
Since the leading behavior of $\Delta D(\omega,L)$ is proportional to $O(e^{ikL})$ we obtain
\begin{align}
  e^{-\omega \tau}\Delta D(\omega, L) &\sim e^{- \left(2m + \frac{k^2}{m}\right) \tau} e^{ikL}
  \sim e^{- \frac{\tau}{m} \left( k - i \frac{mL}{2\tau} \right)^2 - \frac{mL^2}{4\tau} - 2m\tau } 
\end{align}
where we have ignored the dependence of $D(\omega,L)$ on $k$, which is irrelevant in deriving the asymptotic behavior.
Therefore, as $\mu \rightarrow 0$,
\begin{align}
  \Delta C (\tau,L) &= - \int \frac{d\omega}{2\pi i} e^{-\omega \tau} \Delta D(\omega,L) \nonumber\\
  &\sim \int kdk e^{- \frac{\tau}{m} \left( k - i \frac{mL}{2\tau} \right)^2 - \frac{mL^2}{4\tau} - 2m\tau } \nonumber\\
  &\sim e^{ - \frac{1}{4\mu} - 2m\tau }
\end{align}
and the finite-volume imaginary-time correlation function, $C (\tau,L)$, approaches the infinite-volume imaginary-time correlation function, $C (\tau)$ as  
\begin{align}
  C (\tau,L) = C (\tau) \left( 1 + O(e^{ - \frac{1}{4\mu} - 2m\tau }) \right)\rightarrow C (\tau).
\end{align}
\par
Fig.\,{\ref{C_tauL} shows the domains of the asymptotic behavior of the imaginary-time correlation function in the $\tau L$ plane.
$\mu \rightarrow 0$ is the infinite-volume-limit domain and $\mu \rightarrow \infty$ is the lowest-mode-dominance domain.
\section{Quantitative results with simple models}
Now we demonstrate the results described in the previous section using simple nonrelativistic models with a contact interaction, effective field theory  (EFT) \cite{Kaplan:1996xu}, and a separable potential \cite{Yamaguchi:1954mp}.
\subsection{EFT with a contact interaction}
\subsubsection{Correlation function in a finite volume}
The Lagrangiian of EFT of contact interaction is 
\begin{align}
	{\cal L} = \psi^\dagger \left(i\frac{\partial}{\partial t} - m + \frac{1}{2m} \boldsymbol{\nabla}^2 \right)\psi
	-\frac{1}{2} V \psi^\dagger\psi^\dagger\psi\psi,
\end{align}
and the interpolating operators are
\begin{align}
\begin{split}
	{\cal O}(\tau) &= \frac{1}{2L^3} \int d^3x \psi(\tau,\boldsymbol x)\psi(\tau,\boldsymbol x), \\
	{\cal O}^{\dagger}(0) &= \frac{1}{2L^3} \int d^3x \psi^\dagger(0,\boldsymbol x)\psi^\dagger(0,\boldsymbol x).
\end{split}
\end{align}
\par
In the infinite volume the correlation function as a function of the complex energy, $\omega$, on the physical (unphysical) sheet, $D_1(\omega)$ ($D_2(\omega)$), is given as (Fig.\,\ref{D_diag})
\begin{figure}[b]
\begin{center}
\begin{tikzpicture} [thick]
  \draw (-0.5,0) arc (-135:-45:0.70710678);
  \draw (0.5,0) arc (45:135:0.70710678);
  \draw (0,0) arc (-135:-45:0.70710678);
  \draw (1,0) arc (45:135:0.70710678);
  \filldraw[white] (0.25,0) circle [radius=0.375];
  \filldraw[pattern=north east lines] (0.25,0) circle [radius=0.375];
  \draw [red, very thick] (-0.6,0.1) -- (-0.4,-0.1);
  \draw [red, very thick] (-0.6,-0.1) -- (-0.4,0.1);
  \draw [red, very thick] (0.9,0.1) -- (1.1,-0.1);
  \draw [red, very thick] (0.9,-0.1) -- (1.1,0.1);
  \draw (1.5,0) node {$=$}; 
  \draw (2,0) arc (-135:-45:0.70710678);
  \draw (3,0) arc (45:135:0.70710678);
  \draw [red, very thick] (1.9,0.1) -- (2.1,-0.1);
  \draw [red, very thick] (1.9,-0.1) -- (2.1,0.1);
  \draw [red, very thick] (2.9,0.1) -- (3.1,-0.1);
  \draw [red, very thick] (2.9,-0.1) -- (3.1,0.1);
  \draw (3.5,0) node {$+$}; 
  \draw (4,0) arc (-135:-45:0.70710678);
  \draw (5,0) arc (45:135:0.70710678);
  \draw (5,0) arc (-135:-45:0.70710678);
  \draw (6,0) arc (45:135:0.70710678);
  \draw [fill] (5,0) circle [radius=0.05];
  \draw [red, very thick] (3.9,0.1) -- (4.1,-0.1);
  \draw [red, very thick] (3.9,-0.1) -- (4.1,0.1);
  \draw [red, very thick] (5.9,0.1) -- (6.1,-0.1);
  \draw [red, very thick] (5.9,-0.1) -- (6.1,0.1);
  \draw (6.5,0) node {$+$}; 
  \draw (7,0) node [right] {$\cdots$}; 
  \draw (0.25,-0.75) node {$D$}; 
  \draw (2.5,-0.75) node {$G$}; 
  \draw (4.5,-0.75) node {$G$}; 
  \draw (5,-0.75) node {$V$}; 
  \draw (5.5,-0.75) node {$G$}; 
\end{tikzpicture}
\end{center}
\caption{Diagrammatic representation of the correlation function, Eq.\,\eqref{D_inf} and Eq.\,\eqref{D_fin}.}
\label{D_diag}
\end{figure}
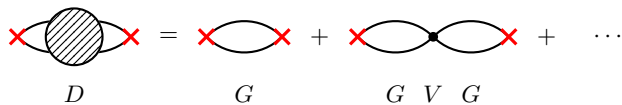
\begin{align}
	D_{1,2}(\omega) &= \frac{G_{1,2}(\omega)}{1 - V G_{1,2}(\omega)},
\label{D_inf}
\end{align}
where $G_1(\omega)$  $(G_2(\omega))$ is the two-particle loop function on the physical (unphysical) sheet,
\begin{align}
	G_{1,2}(\omega)
	&= \pm \int^\Lambda \frac{d^3q}{\left(2\pi\right)^3}\frac{1}{\omega-2m-q^2/m}
	= \pm \frac{m\sqrt{-m(\omega-2m)}}{4\pi} = -i\frac{mk}{4\pi}.
\label{eq:g_inf}
\end{align}
The integral is dimensionally regularized but other regularizations do not change the results.
Then, 
\begin{eqnarray}
  D_{1,2}(\omega) &= \frac{-i\frac{mk}{4\pi}}{1-V\left(-i\frac{mk}{4\pi}\right)} = - \frac{k}{V\left(k-i\frac{4\pi}{mV}\right)} = -\frac{1}{V} - \frac{k_p/V}{k-k_p},
\end{eqnarray}
where $k_p=i\frac{4\pi}{mV}$. It is a peculiar feature of the model that the correlation function is given by a single pole term at $k=k_p$.
If $V>0$ then ${\rm Im}\,k_{p} > 0$ and the pole is a bound-state pole, while if $V<0$ then ${\rm Im}\,k_{p} < 0$ and the pole is a virtual-state pole.
\par
In a finite volume we obtain the correlation function on the physical sheet of the complex energy, $D_1(\omega,L)$, by replacing the integral by a sum in the two-particle loop function as
\begin{align}
	&\int^\Lambda \frac{d^3q}{\left(2\pi\right)^3}\frac{1}{\omega-2m-q^2/m} \longrightarrow \frac{1}{L^3} \sum_{\boldsymbol n}^\Lambda \frac{1}{\omega-2m-q_{\boldsymbol n}^2/m},
\end{align}
where $q_{\boldsymbol n}=2\pi{\boldsymbol n}/L$.
Then, we define the correlation function on the unphysical sheet of the complex energy, $D_2(\omega,L)$, by requiring the same relations as in the infinite volume (Fig.\ref{D_diag}).
Namely,
\begin{align}
	D_{1,2}(\omega,L) &= \frac{G_{1,2}(\omega,L)}{1 - V G_{1,2}(\omega,L)},
\label{D_fin}
\end{align}
where $G_1(\omega,L)$ and $G_2(\omega,L)$ are the two-particle loop functions in a finite volume on the physical sheet and the unphysical sheet, respectively,
\begin{align}
	G_{1,2}(\omega,L)
	&= \pm \frac{1}{L^3} \sum_{\boldsymbol n}^\Lambda \frac{1}{\omega-2m-q_{\boldsymbol n}^2/m}.
\end{align}
\subsubsection{Useful expression of $G_{1,2}(\omega,L)$}
We will derive an useful expression of $G_{1,2}(\omega,L)$ for non-zero ${\rm Im}\,k$. 
$G_{1,2}(\omega,L)$ is given as
\begin{align}
    G_{1,2}(\omega,L)
    &=\mp\frac{m}{4\pi^2L}\lim_{{\hat \Lambda}\to\infty}\biggl[\sum_{\boldsymbol{n}\in\mathbb{Z}^3}^{n<{\hat \Lambda}} \frac{1}{n^2-\hat{k}^2}-4\pi{\hat \Lambda}\biggr],
\label{eq:gxyz_def}
\end{align}
where $\hat{k}=kL/2\pi$ and $\omega=2m+k^2/m=2m+{4\pi}^2{\hat k}^2/(mL^2)$.
Using the Poisson summation formula, we can rewrite the limit in Eq.\,\eqref{eq:gxyz_def} as 
\begin{align}
    \lim_{{\hat \Lambda}\to\infty}\biggl[\sum_{\boldsymbol{n}\in\mathbb{Z}^3}^{n<{\hat \Lambda}} \frac{1}{n^2-\hat{k}^2}-4\pi{\hat \Lambda}\biggr]
    &=\lim_{{\hat \Lambda}\to\infty}\biggl[\int \frac{d^3\chi~\Theta({\hat \Lambda}-\chi)}{\chi^2-\hat{k}^2}\sum_{\boldsymbol{\nu}\in\mathbb{Z}^3}^{}e^{-i2\pi\boldsymbol{\chi}\cdot\boldsymbol{\nu}}-4\pi{\hat \Lambda}\biggr]\nonumber\\
    &=4\pi\hat{k}^2 \int_0^\infty \frac{d\chi}{\chi^2-\hat{k}^2}+\sum_{\boldsymbol{\nu}\in\mathbb{Z}^3, \boldsymbol{\nu}\neq 0}^{}4\pi \int_0^\infty d\chi \frac{\chi^2}{\chi^2-\hat{k}^2}j_0(2\pi \nu \chi)\nonumber\\
    &= 
    \begin{cases}
        \displaystyle 2i\pi^2 \hat{k} + \sum_{\boldsymbol{\nu}\in\mathbb{Z}^3, \boldsymbol{\nu}\neq 0}^{}~\frac{\pi}{\nu}e^{+i2\pi \nu \hat{k}} & (\rm{Im}\,\hat{k} > 0),\\[10pt]
        \displaystyle -2i\pi^2 \hat{k} + \sum_{\boldsymbol{\nu}\in\mathbb{Z}^3, \boldsymbol{\nu}\neq 0}^{}~\frac{\pi}{\nu}e^{-i2\pi \nu \hat{k}} & (\rm{Im}\,\hat{k} < 0),
    \end{cases}\label{eq:gxyz_exp1}
\end{align}
where $\Theta(z)$ is the Heaviside step function and $j_n(z)$ is the spherical Bessel function of the first kind.
Note that we used the analytical expressions for the integrals
\begin{align}
    4\pi\hat{k}^2 \int_0^\infty \frac{d\chi}{\chi^2-\hat{k}^2}
    &=
    \begin{cases}
        +2i\pi^2 \hat{k} & (\rm{Im}\,\hat{k} > 0)\\
        -2i\pi^2 \hat{k} & (\rm{Im}\,\hat{k} < 0)
    \end{cases}
,
\end{align}
and 
\begin{align}
    4\pi \int_0^\infty d\chi \frac{\chi^2}{\chi^2-\hat{k}^2}j_0(2\pi \nu \chi)
    &=
    \begin{cases}
        \displaystyle \frac{\pi}{\nu}e^{+i2\pi \nu \hat{k}} & (\rm{Im}\,\hat{k} > 0)\\[10pt]
        \displaystyle \frac{\pi}{\nu}e^{-i2\pi \nu \hat{k}} & (\rm{Im}\,\hat{k} < 0)
    \end{cases}
.
\end{align}
Substituting Eq.\,\eqref{eq:gxyz_exp1} to Eq.\,\eqref{eq:gxyz_def} we obtain
\begin{eqnarray}
  G_{1,2}(\omega,L)  =  - i \frac{m}{4\pi}\biggl[ k \mp i \frac{1}{L}
    \sum_{\boldsymbol{\nu}\in\mathbb{Z}^3, \boldsymbol{\nu} \ne 0}\frac{1}{\nu}e^{\pm i \nu L k}\biggr].\label{eq:gxyz_exp2}
\end{eqnarray}
Note, the imaginary part of $k$ is ${\rm Im}\,k > 0$ for the physical sheet and ${\rm Im}\,k< 0$ for the unphysical sheet.
And thus, this expression has good convergence on both sheets (when $k$ has a non-zero imaginary part).
\subsubsection{Pole energies in a finite volume}
Having defined the correlation function on the unphysical sheet in a finite volume we discuss here the volume dependences of pole energies on the unphysical sheet as well as the physical sheet.
The volume dependences of pole energies on the physical sheet are known and we will check if known results are reproduced, while the volume dependences of poles on the unphysical sheet are unknown and we will obtain new results.
\par
\paragraph{Bound/Virtual State}
Next, we derive the analytical expression of the finite volume correction to the energies of the bound and virtual states, $\omega_b(L)$ and $\omega_v(L)$, which satisfy
\begin{align}
  1 - V G_{1,2}(\omega_{b,v}(L),L) = 0. 
\label{eq:pole_bv}
\end{align}
Substituting Eq.\,\eqref{eq:gxyz_exp2} into Eq.\,\eqref{eq:pole_bv} and denoting $\omega_{b,v}(L) = 2m + k_{b,v}(L)^2/m$ we obtain
\begin{align}
  i \frac{4\pi}{mV} & = k_{b,v}(L) \mp
   i \frac{1}{L}
   \sum_{\boldsymbol{\nu}\in\mathbb{Z}^3, \boldsymbol{\nu} \ne 0}\frac{1}{\nu}e^{\pm i \nu L k_{b,v}(L)} \nonumber\\
   & =
   k_{b,v}(L) \mp i \biggl(\frac{6}{L}e^{\pm i L k_{b,v}(L)} +
   \frac{12}{\sqrt{2}L}e^{\pm i \sqrt{2} L k_{b,v}(L)} + \cdots \biggr).
\end{align}
Then, $k_{b,v}(L)$ is solved as
\begin{eqnarray}
  k_{b,v}(L) = \pm i \kappa \biggl[ 1 + \frac{6}{\kappa L}e^{- L \kappa} + O(e^{- \sqrt{2} \kappa L})\biggr],
\end{eqnarray}
where $\kappa= 4\pi/(m |V|)$ and only the leading-order correction is explicitly shown.
Therefore, the finite volume correction to the energy of the virtual state exponentially decreases
as the volume size increases just like that of the bound state as
\begin{eqnarray}
  \omega_{b,v}(L) = \omega_{b,v} + \frac{12 \kappa}{m L}e^{-\kappa L} + O(e^{- \sqrt{2} \kappa L}).
\label{eq:pole_asymp_bv}
\end{eqnarray}
Here $\omega_{b,v} = 2m - \kappa^2/m$, is the energy of the bound (virtual) state in the infinite volume.
\paragraph{Scattering States}
In a finite volume the energies of continuous scattering states in the infinite volume become discrete.
Each discrete energy is related to the scattering phase shift at the energy, the L\"{ui}scher equation \cite{Luscher:1990ux},
which can be easily demonstrated for the present model as done in Ref.\,\cite{Doring:2011vk}.

In the infinite volume the scattering $T$-matrix is given by
\begin{align}
   T(\omega) = \frac{V}{1-VG_1(\omega)}.
\end{align}
In a finite volume the discrete energy, $\omega_s(L)$, satisfies
\begin{align}
   1-VG_1(\omega_s(L),L) = 0
\end{align}
Therefore, at the energy, $\omega_s(L)$, the $T$-matrix can be rewritten as
\begin{align}
   T(\omega_s(L))  = \frac{V}{1-VG_1(\omega_s(L))} = \frac{1}{G_1(\omega_s(L),L)-G_1(\omega_s(L))}
\end{align}
Since $T(\omega)$ is related to the scattering phase shift, $\delta(k)$, as
\begin{align}
   T(\omega) = \frac{-4\pi / m}{k \cot\delta(k)-ik}
\end{align}
we obtain
\begin{align}
   k_s(L) \cot\delta(k_s(L)) = - \frac{4\pi}{m} \left[ G_1(\omega_s(L),L) - G_1(\omega_s(L)) \right] + ik_s(L)
\label{eq:kcotd}
\end{align}
where $\omega_s(L) = 2m + k_s(L)^2/m$.
Substituting Eqs.\,\eqref{eq:g_inf}, \eqref{eq:gxyz_def} into Eq.\,\eqref{eq:kcotd} and noting that
\begin{align}
   \lim_{\hat\Lambda \to \infty} \left[ \sum_{\boldsymbol{n}\in\mathbb{Z}^3}^{n < \hat \Lambda} \frac{1}{{\boldsymbol n}^2- \hat{k}^2} - 4\pi \hat\Lambda\right] = 2\pi^{1/2} {\cal Z}_{00}(1,\hat k^2)
\end{align}
where ${\cal Z}_{00}(1,\hat k^2)$ is the L\"{ui}scher zeta-function, we reproduce the L\"{ui}scher equation \cite{Luscher:1990ux}
\begin{align}
   k_s(L) \cot\delta(k_s(L)) = \frac{2\pi}{L}\pi^{-3/2} {\cal Z}_{00}(1,\hat k_s(L)^2).
\label{eq:Luscher}
\end{align}

In Ref.\,\cite{Luscher:1986pf} the $1/L$ expansion of the energy shift of the lowest scattering state in a finite volume is given.
It is interesting to see how the $1/L$ expansion is carried out in the present model, which is a typical example of the degenerate perturbation expansion.
In order to expand the energy of the lowest scattering state, $\omega_s(L)$, in $1/L$ we rewrite $G_1(\omega,L)$ as
\begin{align}
   G_1(\omega,L)
	&= \frac{1}{L^3(\omega-2m)} - \frac{m}{4\pi^2 L} \lim_{\hat\Lambda \to \infty} \left[ \sum_{\boldsymbol{n}\in\mathbb{Z}^3, \boldsymbol{n}\ne 0}^{n < \hat \Lambda} \frac{1}{{\boldsymbol n}^2- \hat{k}^2} - 4\pi \hat\Lambda\right] \nonumber\\
	&\equiv \frac{1}{L^3(\omega-2m)} + \tilde G_1(\omega,L),
\end{align}
and then $D_1(\omega,L)$ as
\begin{align}
	D_1(\omega,L)
	=& \frac{V}{1 - V \tilde G_1(\omega,L)} + \frac{V}{1 - V \tilde G_1(\omega,L)} \frac{1}{L^3 (\omega - 2m) - \frac{V}{1 - V \tilde G_1(\omega,L)}}   \frac{V}{1 - V \tilde G_1(\omega,L)}.
\end{align}
$\omega_s(L)$ satisfies
\begin{align}
	\omega_s(L) - 2m = \frac{1}{L^3} \frac{V}{1-V \tilde G_1(\omega_s(L),L) }.
\end{align}
The expansion of $\omega_s(L)$ in $L$ is obtained as
\begin{align}
	\omega_s(L) - 2m &= \frac{1}{L^3} \left[ V + \tilde G_1(2m,L) V^2 + \tilde G_1(2m,L)^2 V^3 + \frac{V}{L^3} \left.\frac{d\tilde G_1(\omega,L)}{d\omega}\right|_{\omega=2m} V^2 + \cdots \right] \nonumber \\
	&= \frac{1}{L^3} \left[ V - \frac{m}{4\pi^2L} Z_{00}(1,0) V^2 + \frac{m^2}{16\pi^4L^2} Z_{00}(1,0) V^3 - \frac{m^2L}{16\pi^4} Z_{00}(2,0) \frac{V^3}{L^3}  + \cdots \right] \nonumber \\
	&= \frac{1}{L^3} \frac{4\pi a}{m} \left[ 1 - \frac{a}{\pi L} Z_{00}(1,0) + \frac{a^2}{\pi^2 L^2} \left( Z_{00}(1,0)^2 - Z_{00}(2,0) \right)  + \cdots \right],
\label{eq:omega_sc_exp}
\end{align}
 where $a=-\frac{M}{4\pi}V$ is the scattering length, $Z_{00}(1,0)$ and $Z_{00}(2,0)$ are subtracted L\"{u}scher Zeta functions,
\begin{align}
	Z_{00}(1,\hat k^2) &= \lim_{\hat\Lambda \to \infty} \left[ \sum_{\boldsymbol{n}\in\mathbb{Z}^3, \boldsymbol{n}\ne 0}^{n < \hat \Lambda} \frac{1}{{\boldsymbol n}^2- \hat k^2} - 4\pi \hat\Lambda\right], \\
	Z_{00}(2,\hat k^2) &= \sum_{\boldsymbol{n}\in\mathbb{Z}^3, \boldsymbol{n}\ne 0}^{n < \hat \Lambda} \frac{1}{({\boldsymbol n}^2- \hat k^2)^2}.
\end{align}
Eq.\,\eqref{eq:omega_sc_exp} is nothing but the $1/L$ expansion of the energy shift of the lowest scattering state given in \cite{Luscher:1986pf},
\begin{align}
	\omega_s(L) = 2m - \frac{4\pi a}{mL^3}\left\{1+c_1\frac{a}{L}+c_2\frac{a^2}{L^2}\right\} + O(L^{-6}),
\label{eq:Luscher_exp}
\end{align}
where $c_1 = \frac{1}{\pi} Z_{00}(1,0)=-2.837293$ and $c_2 = \displaystyle\frac{1}{\pi^2} \left( Z_{00}(1,0)^2 - Z_{00}(2,0) \right)=6.375183$.
\par
Fig.\ref{pole_fin} shows numerical results of the volume dependence of pole energies for the bound state, the virtual state and the lowest scattering state, where the parameters are taken to be $m=1$ and $V=1/(0.16)^2$ in arbitrary units.
The numerical results are consistent with the asymptotic formulas for large volume, $12\kappa_{b,v} \exp(-\kappa_{b,v} L)/mL$ and Eq.\,\eqref{eq:Luscher_exp}, respectively.
\begin{figure}[h]
\begin{minipage}{0.495\columnwidth}
	\includegraphics[width=\linewidth]{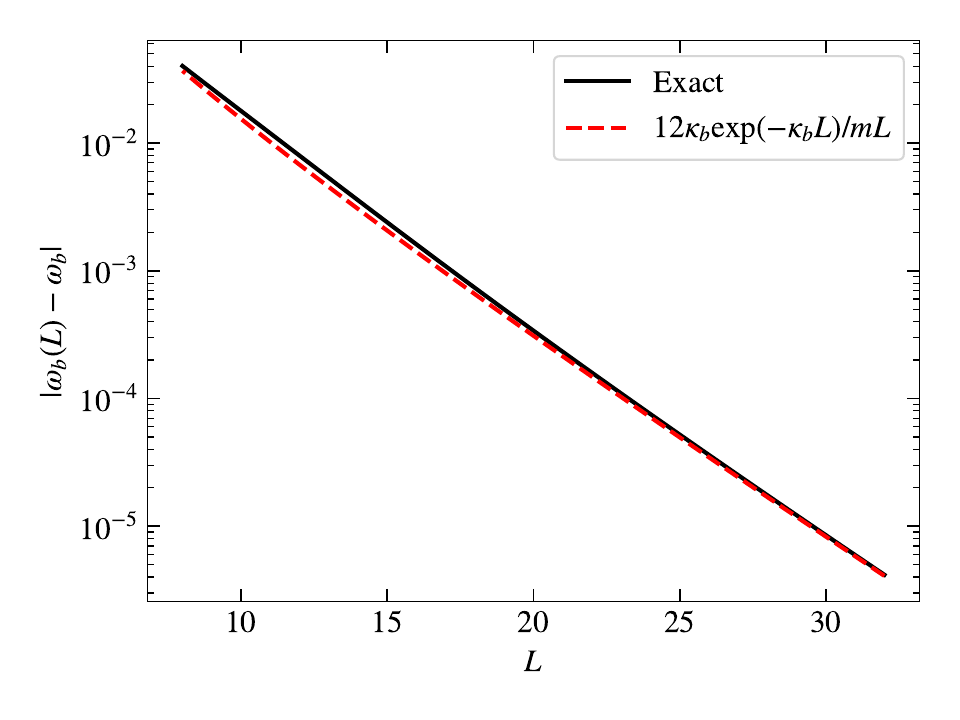} 
\end{minipage}
\begin{minipage}{0.495\columnwidth}
	\includegraphics[width=\linewidth]{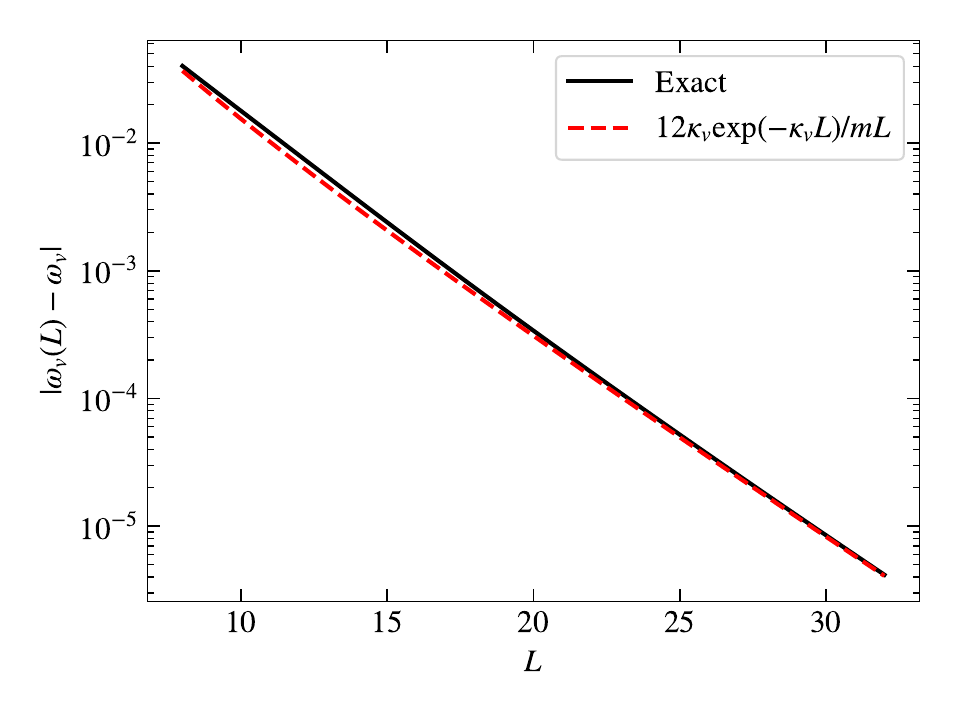} 
\end{minipage}
\begin{minipage}{0.495\columnwidth}
	\includegraphics[width=\linewidth]{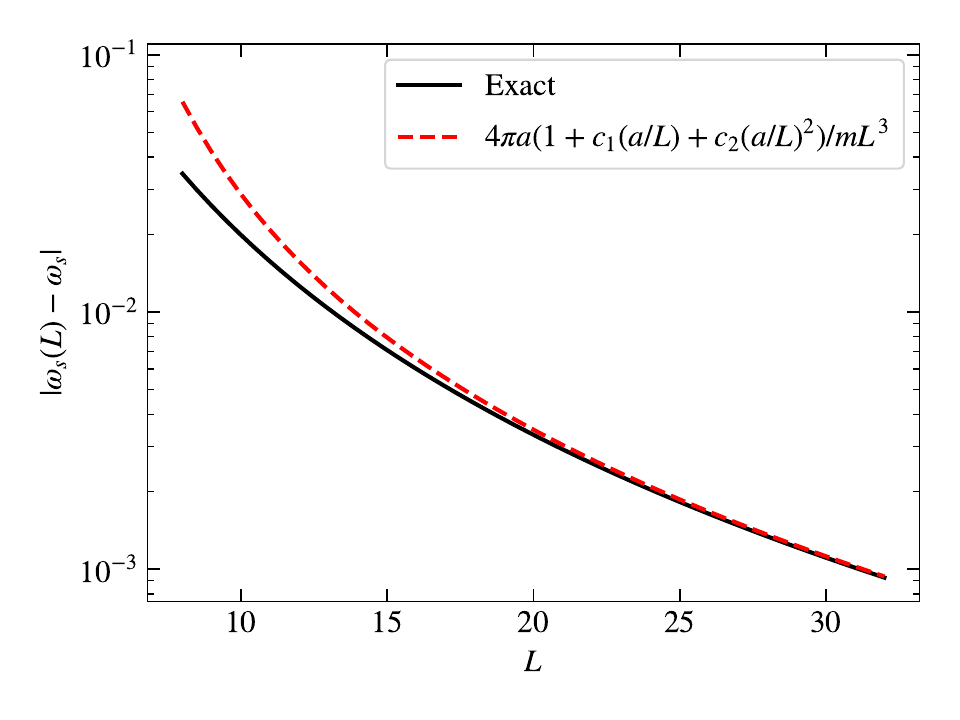} 
\end{minipage}
\caption{Volume dependence of the finite-volume correction of the pole energies for the EFT for the bound state (top left), the virtual state (top right) and the lowest \lq\lq scattering" state (bottom).
\lq\lq Exact" means exact results within the model, Eq.\,\eqref{eq:pole_bv}, for bound and virtual states and Eq.\,\eqref{eq:Luscher} for the lowest \lq\lq scattering" state. The asymptotic formulas for large volume, Eq.\,\eqref{eq:pole_asymp_bv}, for bound and virtual states and Eq.\,\eqref{eq:Luscher_exp}, for the lowest \lq\lq scattering" state are also shown for comparison.}
\label{pole_fin}
\end{figure}

\subsubsection{Finite volume correction of imaginary-time correlation function}
Now, we show the imaginary-time correlation function in a finite volume, $C(\tau,L)$, calculated using EFT with a contact interaction.
Fig.\,\ref{C_tau_and_C_L} shows, $C(\tau,L)$, vs. $\tau$, for a fixed volume, $L=8$, (left) and  $C(\tau,L)$, vs. $L$ for a fixed imaginary time, $\tau=12$, (right) where two domains, the infinite-volume domain and the lowest-mode-dominance domain and the transition between them are clearly observed.
\begin{figure}[h]
\begin{minipage}{0.495\columnwidth}
	\includegraphics[width=\linewidth]{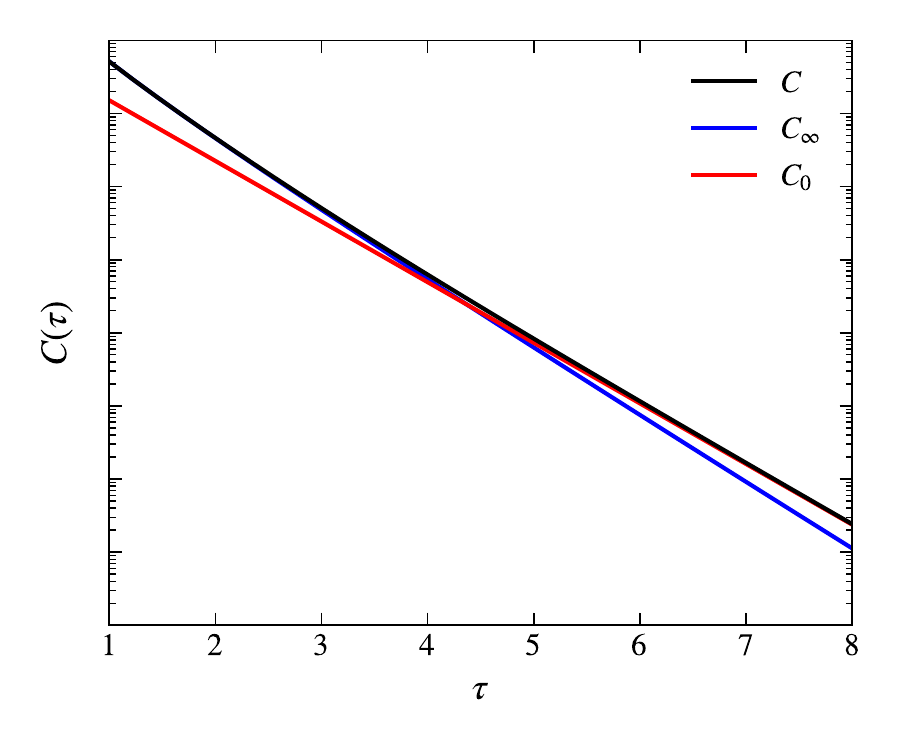} 
\end{minipage}
\begin{minipage}{0.495\columnwidth}
	\includegraphics[width=\linewidth]{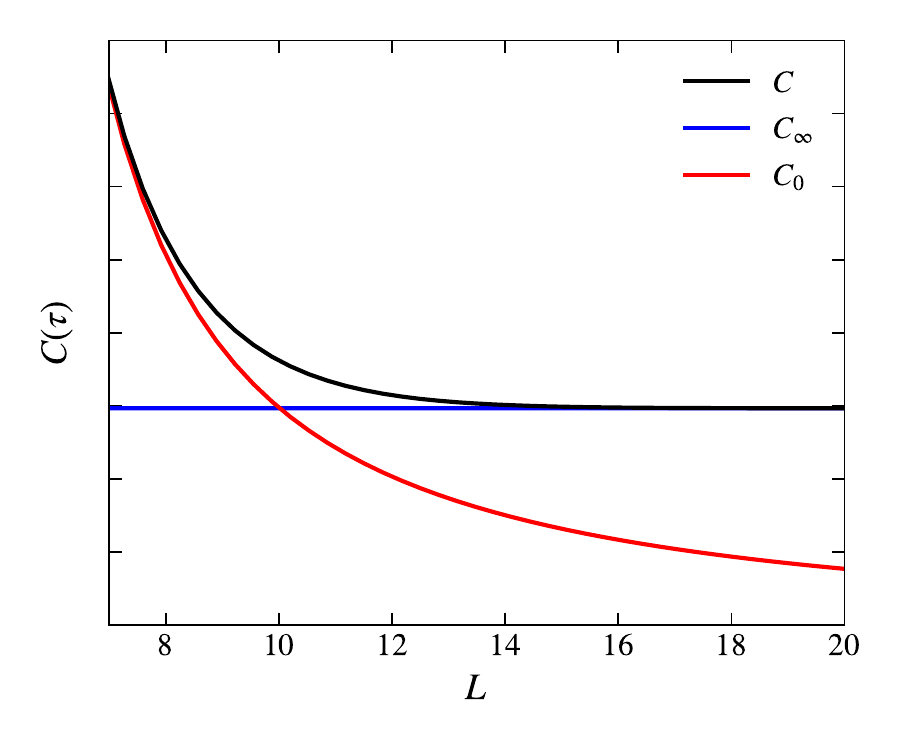} 
\end{minipage}
\caption{The imaginary-time correlation function vs. imaginary time $\tau$ with a fixed volume, $L=8$, (left) and the imaginary-time correlation function vs. the size of the volume, $L$ with fixed time, $\tau=12$ (right).
The black, blue and red lines represent the imaginary-time correlation function in a finite volume, $C$, in the infinite volume, $C_\infty$, and the lowest-mode contribution, $C_0$, respectively.}
\label{C_tau_and_C_L}
\end{figure}

We then discuss the finite volume correction of the imaginary-time correlation function in the small $\mu$ region in detail.
}
On the right-hand-side of Eq.\,\eqref{eq:gxyz_exp2} the first term is the loop function in the infinite volume and the rest is the finite-volume correction,
\begin{align}
   G_{1,2}(\omega,L) &= G_{1,2}(\omega) + \Delta G_{1,2}(\omega,L),
\end{align}
where
\begin{align}
  \Delta G_{1,2}(\omega, L) = \mp \frac{m}{4\pi L} \sum_{\boldsymbol{\nu} \ne 0}\frac{1}{\nu}e^{i L \nu k}.
\end{align}
When $L\,{\rm Im}k \gg 1$, $\Delta G_{1,2}(\omega, L)$ is dominated by the terms of $\nu=1$ and behaves as
\begin{align}
  \Delta G_{1,2}(\omega, L) \sim \mp \frac{3m}{2\pi L} e^{i L k}.
\end{align}
The finite-volume correction, $\Delta D_{1,2}(\omega,L) = D_{1,2}(\omega,L) - D_{1,2}(\omega)$, is given by
\begin{align}
	\Delta D_{1,2}(\omega,L) &= \frac{\Delta G_{1,2}(\omega,L)}{\left[ 1 - V G_{1,2}(\omega) \right] \left[ 1 - V \left(G_{1,2}(\omega) + \Delta G_{1,2}(\omega,L)\right) \right]} \nonumber \\
	&= \frac{\Delta G_{1,2}(\omega,L)}{\left[ 1 - V G_{1,2}(\omega) \right]^2} + \cdots
\end{align}
and behaves as
\begin{align}
	\Delta D_{1,2}(\omega,L) &\sim \mp \frac{3m}{2\pi L} e^{i L k} \frac{k_p^2}{\left( k - k_p \right)^2}.
\end{align}
Then, the finite-volume correction, $\Delta C(\omega, L) = C(\omega, L) - C(\omega)$, is given by
\begin{align}
  \Delta C(\omega, L)
  &= - \frac{1}{2\pi i} \int d\omega e^{-\omega\tau} \frac{\Delta G_1(\omega,L)}{\left[ 1 - V G_1(\omega) \right]^2} + \cdots \nonumber \\
  &\sim - \frac{1}{2\pi i} \int d\omega e^{-\omega\tau} \left( - \frac{3m}{2\pi L} e^{i L k} \frac{k_p^2}{\left( k - k_p \right)^2} \right).
\end{align}
\begin{figure}[b]
\begin{minipage}{0.495\columnwidth}
	\includegraphics[width=\linewidth]{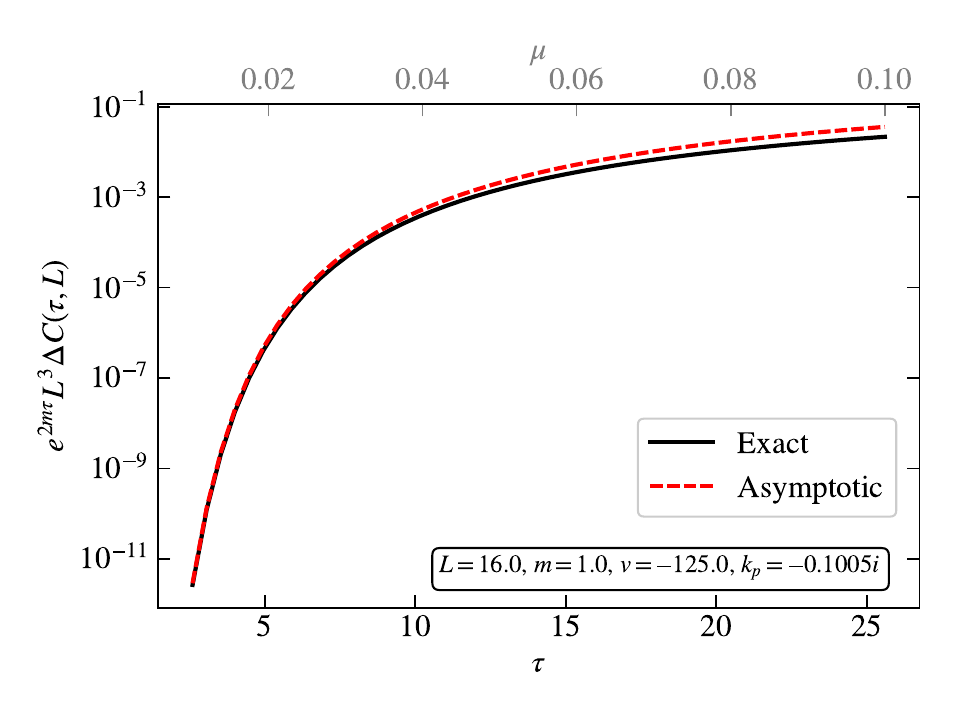} 
\end{minipage}
\begin{minipage}{0.495\columnwidth}
	\includegraphics[width=\linewidth]{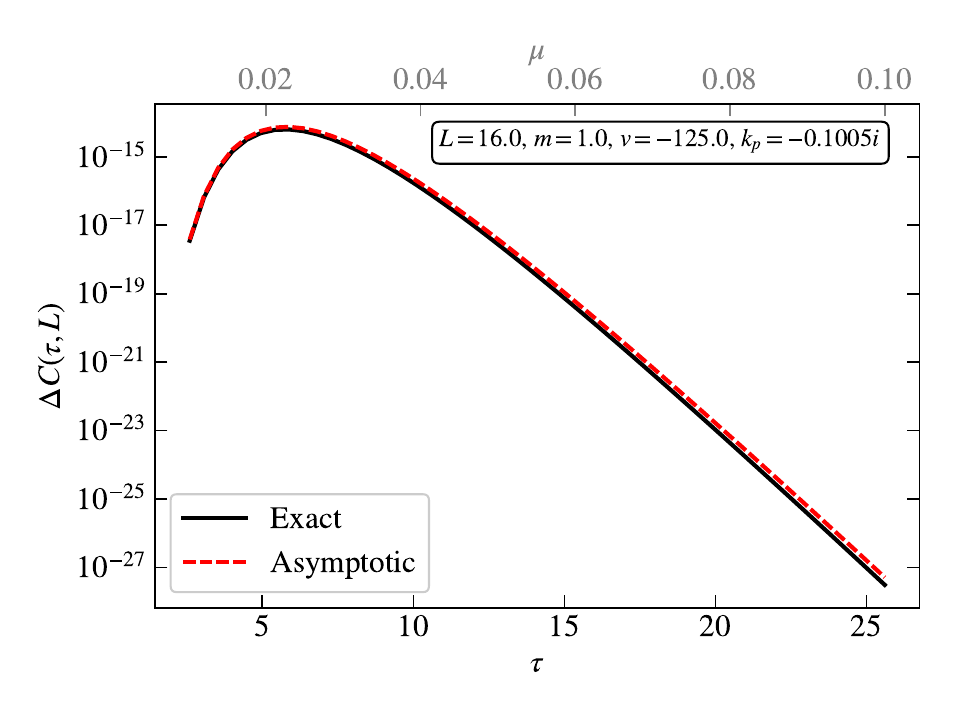} 
\end{minipage}
\caption{$e^{2m\tau}L^3\Delta C(\omega, L)$ (left) and $\Delta C(\omega, L)$ (right) vs. $\tau$ with a fixed size of the volume, $L=16$.}
\label{deltaC_tau}
\begin{minipage}{0.495\columnwidth}
	\includegraphics[width=\linewidth]{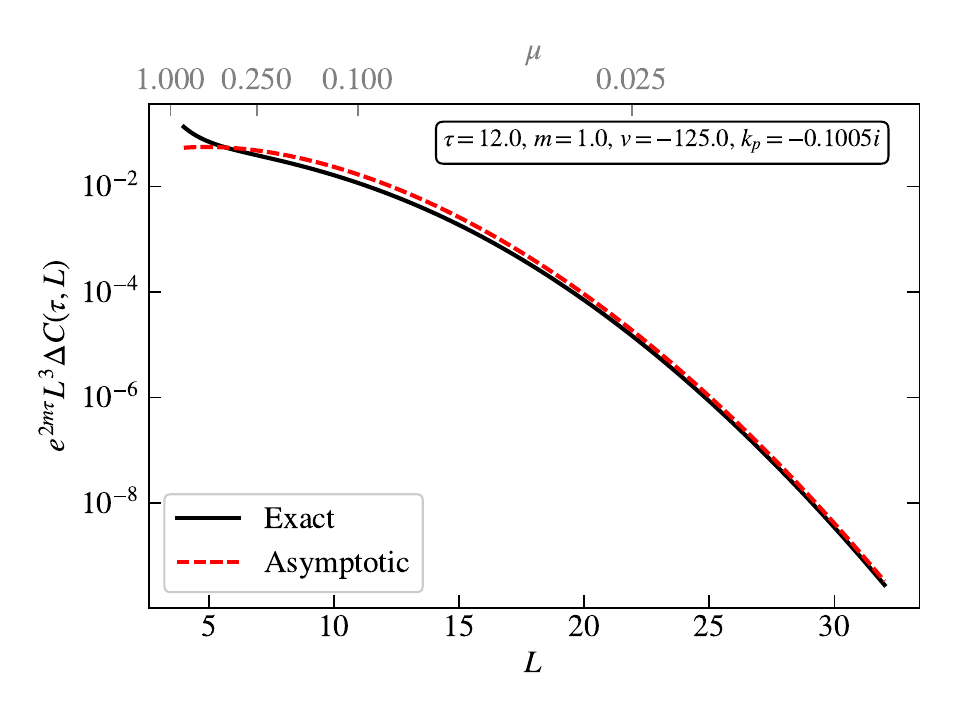} 
\end{minipage}
\begin{minipage}{0.495\columnwidth}
	\includegraphics[width=\linewidth]{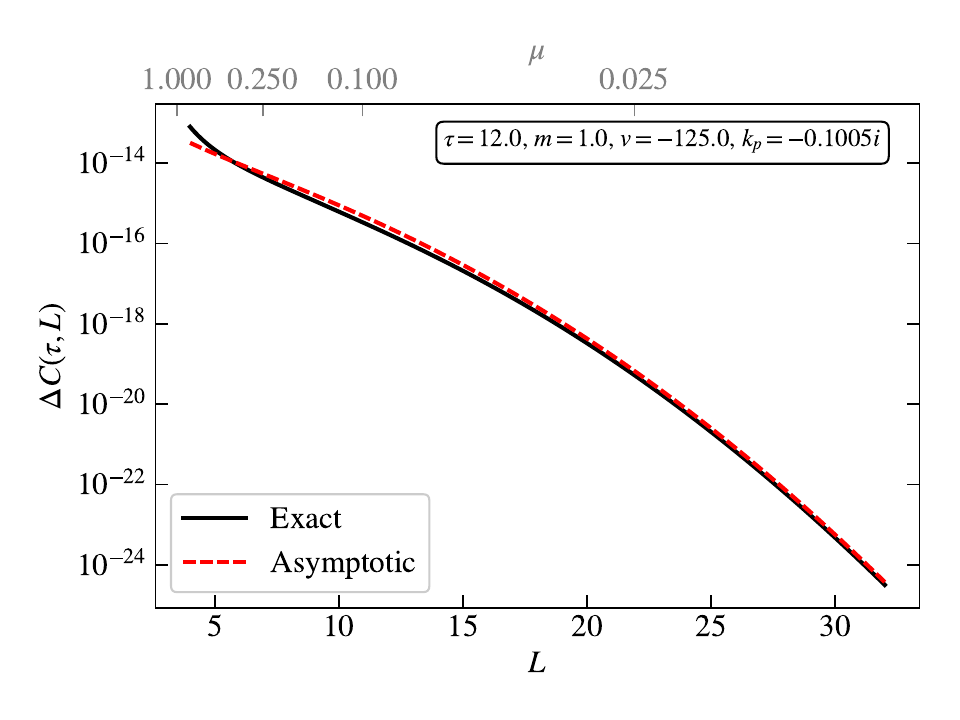} 
\end{minipage}
\caption{$e^{2m\tau}L^3\Delta C(\omega, L)$ (left) and $\Delta C(\omega, L)$ (right) vs. $L$ with a fixed imaginary time, $\tau=12$.}
\label{deltaC_L}
\end{figure}
After taking out $e^{-2m\tau} L^{-3}$, $\Delta C(\omega, L)$ scales with two parameters, $\mu=\tau/mL^2$ and $x_p=\sqrt{\tau/m}k_p$ as 
\begin{align}
  e^{2m\tau} L^3 \Delta C(\omega, L) &\sim \frac{6x_p^2 e^{-1/4\mu}}{2\pi\mu} \int \frac{dx}{2\pi i} x e^{-(x-i/2\mu)^2} \frac{1}{(x-x_p)^2} \nonumber \\
  &\sim \frac{3x_p^2 e^{-1/4\mu}}{2\pi\mu} \left[ -w\left(\frac{i}{2\sqrt{\mu}}-x_p\right) + x_p w'\left(\frac{i}{2\sqrt{\mu}}-x_p\right) \right],
\end{align}
where $w$ is the Faddeeva function,
\begin{align}
  w(z) = e^{-z^2} {\rm erfc}(-iz)
\end{align}
and $w'(z)$ is its derivative.
At the leading order of $1/z$
\begin{align}
  w(z) \sim \frac{i}{\sqrt{\pi}z}, \quad w'(z) = -\frac{i}{\sqrt{\pi}z^2},
\end{align}
and therefore
\begin{align}
  e^{2m\tau} L^3 \Delta C(\omega, L)
  &\sim \frac{3x_p^2 e^{-1/4\mu}}{2\pi^{3/2}} \left[ -\frac{2}{\sqrt{\mu}} + 8 i x_p + O(\sqrt{\mu}) \right].
\end{align}
In terms of $\omega$ and $L$, $\Delta C(\omega, L)$ behaves asymptotically as
\begin{align}
  \Delta C(\omega, L) \rightarrow
  -\frac{3\kappa_p^2\sqrt{\tau}}{\pi^{3/2}\sqrt{m}L^2} e^{-mL^2/4\tau-2m\tau}.
\end{align}
In Figs.\,\ref{deltaC_tau} and \ref{deltaC_L} we show $\Delta C(\omega, L)$ and its asymptotic form.
\subsection{Finite-volume effect on resonance states in a separable potential model}
Using a separable potential model, we extend our analysis to resonance states and demonstrate that the finite-volume effect on their pole positions is also exponentially suppressed with $L$.
\par
We use the monopole separable potential,
\begin{align}
  \bra{\bm{q}'} \hat V \ket{\bm{q}} = v g(q')g(q),
  \label{eq:sep_pot}
\end{align}
where the center-of-mass degrees of freedom of the two-particle system are taken out, $\bm{q}$ and $\bm{q}'$ are relative momenta and $g(q)  = \beta^2/(\beta^2 + q^2)$.
It should be noted that the contact interaction of EFT in the previous subsection is written as $\bra{\bm{q}'} \hat V \ket{\bm{q}} = V$ in the notation of Eq.\eqref{eq:sep_pot} and in the limit, $\beta \to \infty$, a separable potential model approaches EFT.

The energy of the resonance state $\omega_r= 2m + k_r^2/m$ in the infinite volume is obtained from the condition
\begin{align}
  1 - v G_2^{sep}(\omega_r) = 0,\label{eq:pole_cond_infv}
\end{align}
where $G_2^{sep}(\omega)$ is the analytic continuation of the Green's function on the physical sheet, $G_1^{sep}(\omega)$, and is given as
\begin{align}
  G_2^{sep}(\omega) & = G_1^{sep}(\omega) + i \frac{m k}{2\pi}g^2(k),\\
  G_1^{sep}(\omega) & = \int \frac{d^3q}{(2\pi)^3} \frac{g^2(q)}{\omega - 2m - q^2/m}.      
\end{align}
Here, ${\rm Im}\,k > 0$ for complex $\omega$.
$G^{sep}_{1,2}(\omega)$ can be rewritten in a similar form to the EFT model as,
\begin{align}
  G^{sep}_{1,2}(\omega) = [G_1^{sep}(\omega) - g^2(k) G_1(\omega)] \pm g^2(k) G_1(\omega), \label{eq:gsep12}
\end{align}
with $G_1(\omega)$ in Eq.\,\eqref{eq:g_inf}.
The first term of Eq.\,\eqref{eq:gsep12}, which vanishes in the EFT model, is given by
\begin{align}
  G_1^{sep}(\omega) - g^2(k) G_1(\omega) = 
 m g^2(k)\lim_{\Lambda \rightarrow \infty}\biggl[
     \int^\Lambda \frac{d^3q}{(2\pi)^3}\biggl(
    \frac{1}{q^2 + \beta^2} + \frac{k^2 + \beta^2}{(q^2 + \beta^2)^2}\biggr)
    - \frac{\Lambda}{2\pi^2}\biggr],
\end{align}
in this separable potential model.
This integrand is regular for real $q$, whereas the integrand in the second term of Eq.\,\eqref{eq:gsep12}, $G_1(\omega)$,
is singular at $q=k$ for real $E$.

Now we can translate the above formula to finite volume.
By the replacement
\begin{align}
    d^3q &\to (2\pi/L)^3 \sum_{\bm{n}}, \\
    q^2 &\to (2\pi/L)^2 n^2,
\end{align}
we obtain,
\begin{gather}
    G^{sep}_{1,2}(\omega,L)=-\frac{m}{4\pi^2 L}g^2(k)[-Z_{00}(1,-\hat{\beta}^2)-(\hat{k}^2+\hat{\beta}^2)Z_{00}(2,-\hat{\beta}^2) \pm Z_{00}(1,\hat{k}^2)]
\end{gather}
where $\hat{k}=kL/2\pi$ and $\hat{\beta}=\beta L/2\pi$. 

Here we evaluate the finite-volume correction to the resonance pole position.
The resonance pole condition in a finite volume is
\begin{gather}
  1 - v G_2^{sep}(\omega_r(L),L) = 0,\label{eq:pole_cond_fv}
\end{gather}
where $\omega_r(L)$ is the resonance pole position (energy) in a finite volume.
Using Eqs.\,\eqref{eq:pole_cond_infv}, \eqref{eq:pole_cond_fv}, we obtain,
\begin{gather}
    \omega_r(L) \approx \omega_r + \frac{8\pi k_r (k_r+i\beta)^3}{m^2\beta^3}\Delta G_2^{sep}(\omega_r,L),
\end{gather}
where $\Delta G_2^{sep}(\omega_r,L)=G_2^{sep}(\omega_r,L)-G_2^{sep}(\omega_r)$.
By the use of the Poisson summation formula as in Eq.\,\eqref{eq:gxyz_exp1}, we obtain
\begin{gather}
    \Delta G_2^{sep}(\omega_r,L)=\frac{m}{4\pi^2 L}g^2(k_r)\sum_{\boldsymbol{\nu}\in\mathbb{Z}^3, \boldsymbol{\nu}\ne 0}\biggl[\frac{\pi}{\nu}e^{-\nu\beta L}+(k_r^2+\beta^2)\frac{\pi L}{2\beta}e^{-\nu\beta L}+\frac{\pi}{\nu}e^{-i\nu k_rL}\biggr]
\end{gather}
and hence at the leading order,
\begin{gather}
    \omega_r(L) \simeq \omega_r + \frac{6k_r (k_r+i\beta)^3}{m(k_r^2+\beta^2)^2}\biggl[(k_r^2+\beta^2+\frac{2\beta}{L})e^{-\beta L}+\frac{2\beta}{L} e^{-ik_rL}\biggr]. \label{eq:omgR_Ldependence}
\end{gather}
The finite-volume effect originating from the regular term, $G_1^{sep} - g^2(k)G_1$
is suppressed as $e^{-\beta L}$, while that originating from the singular term $G_1$, behaves as $e^{-i k_r L}$.
In this model, $k_r=\pm\alpha-i\beta$ where $\alpha^2=mv\beta^3/8\pi~(v>0)$, and thus both of them are suppressed by $e^{-\beta L}$.

\begin{figure}[h]
    \centering
    \includegraphics[width=0.7\linewidth]{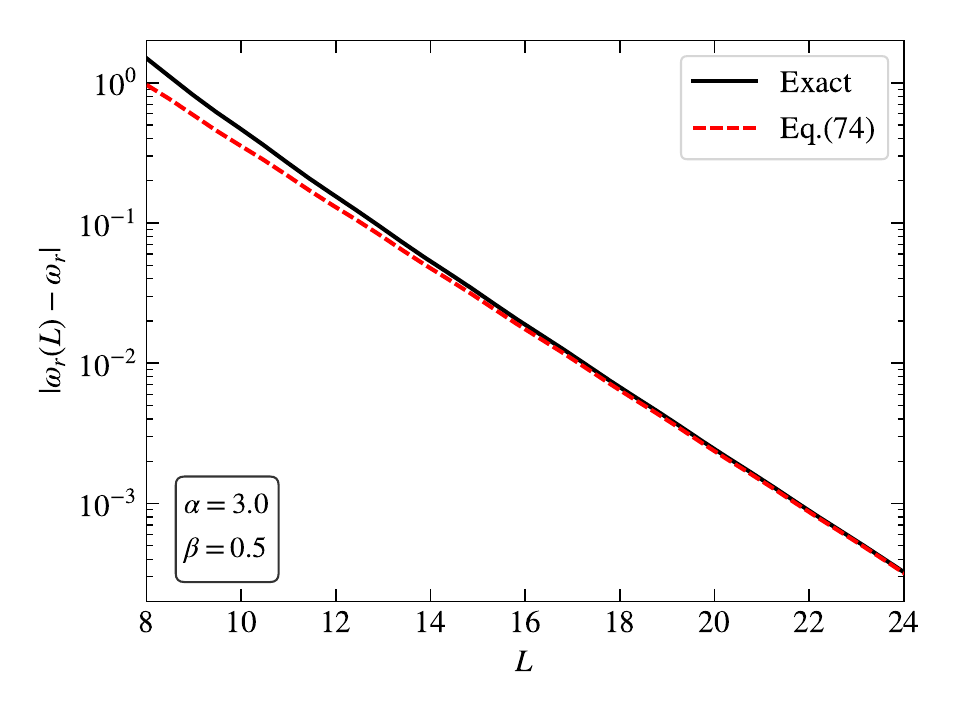}
    \caption{Volume dependence of the finite-volume correction of the resonance pole energy for the separable potential model with parameters $\alpha=3.0$ and $\beta=0.5$.
\lq\lq Exact" means the exact result within the model, Eq.\,\eqref{eq:pole_cond_fv}, and the asymptotic formula, Eq.\,\eqref{eq:omgR_Ldependence}, is also shown for comparison.}
    \label{fig:ep_pole_diff0521}
\end{figure}

We now numerically demonstrate the finite-volume correction to the resonance pole in the separable model.
In Fig.\,\ref{fig:ep_pole_diff0521}, we plot the $L$ dependence of the shift of the resonance pole $|\omega_r(L)-\omega_r|$ for parameters $\alpha=3.0$ and $\beta=0.5$.
We observe that the $L$-dependence obeys Eq.\,\eqref{eq:omgR_Ldependence}.
\section{summary}
We have investigated the two-particle correlation function and unstable states (resonances, virtual states $\cdots$) in a finite volume.
In the Introduction we raised several questions.
The answers we have obtained throughout this paper are as follows.

\begin{itemize}
\item
How do we identify unstable states in a finite volume in general? \\
{\it We define the unphysical sheet (the second sheet in the single-channel case) of the correlation function or the $S$-matrix and identify unstable states as poles on the unphysical sheet.}
\item
How large is the finite volume effect of unstable states, power, $L^{-3}$, or exponential, $\exp(-\# L)$, as the size of the volume, $L$, increases?\\
{\it The finite volume effect of unstable states is exponential, $\exp(-dL)$, where $d$ is determined by the poles of the unstable poles.} \\
The $L$ dependences of pole energies on physical and unphysical sheets in a finite volume are determined by the summation formula
\begin{align*}
   \sum_{\boldsymbol{n}\in\mathbb{Z}^3}^{\Lambda} \frac{1}{{\boldsymbol n}^2- \hat{k}^2} = 
   \begin{cases}
   \displaystyle{ 4 \pi \hat{k}^2 \int_0^\infty \frac{d\chi}{\chi^2-\hat{k}^2} + O(e^{-2\pi L |{\rm Im}\, \hat k|}) } & ({\rm Im}\,\hat k \ne 0) \\
   \displaystyle{ 2\pi^{1/2} {\cal Z}_{00}(1,\hat k^2) } & ({\rm Im}\,\hat k = 0)
   \end{cases}
\end{align*}
and are summarized as follows.
\begin{enumerate}
\item{\lq\lq scattering" state} (physical sheet)
\begin{align*}
	k(L) = k_s(L) \longrightarrow k_s(L) \cot\delta(k_s(L)) = \frac{2\pi}{L}\pi^{-3/2} {\cal Z}_{00}(1,\hat k_s(L)^2)
\end{align*}
\item{bound state} (physical sheet)
\begin{align*}
	k(L) = i\kappa_b(L) \longrightarrow \omega_b(L) = \omega_b  + O(e^{-\kappa_bL})
\end{align*}
\item{virtual state} (unphysical sheet)
\begin{align*}
	k(L) =  - i\kappa_v(L) \longrightarrow \omega_v(L) = \omega_v  + O(e^{-\kappa_vL})
\end{align*}
\item{resonance and anti-resonance} (unphysical sheet)
\begin{align*}
	k(L) = \pm k_r(L) - i \kappa_r(L) \longrightarrow \omega_r(L) = \omega_r  + O(e^{-\kappa_rL})
\end{align*}
\end{enumerate}
The $L$ dependences of pole energies on the physical sheet, 1 and 2, have already been known while those on the unphysical sheet, 3 and 4, are newly obtained in the present paper.  
\item
How does the finite-volume imaginary-time correlation function behave asymptotically? \\
{\it The imaginary-time correlation function of a finite volume approaches that of the infinite volume as $\mu \rightarrow 0$
\begin{align*}
  C (\tau,L) = C (\tau) \left( 1 + O(e^{ - \frac{1}{4\mu} - 2m\tau }) \right)\rightarrow C (\tau),
\end{align*}
while it is dominated by the lowest energy mode as $\mu \rightarrow \infty$, where $\mu=\tau/(m L^2)$, $\tau$ is the imaginary time, $L$ is the size of the volume, and $m$ is the hadrom mass.
\begin{align*}
  C (\tau,L) = \rho(0,L) e^{-\omega(0,L) \tau} \left( 1 + O(e^{-4\pi^2 \mu}) \right)
  \rightarrow \rho(0,L) e^{-\omega(0,L) \tau}.
\end{align*}
}
\item
How can one extract information of unstable states in the infinite volume such as the mass and the width from the imaginary-time correlation function obtained by lattice simulations in a finite volume?\\
{\it One can extract information of unstable states in the infinite volume such as the mass and the width by fitting the form of the pole expansion in terms of the uniformization variable to the imaginary-time correlation function obtained by lattice simulations in a finite volume in the region, $\mu \ll 1$, where the infinite-volume approximation is valid.}
\end{itemize}
These results are demonstrated in the single-channel with a simple interaction of EFT and a separable-potential model. However the results are expected to hold in cases of coupled-channels and/or with more complicated interactions because the loop function plays an essential role which holds true.
\par
While we are preparing the present paper we became aware of Ref.\,\cite{Meissner:2010rq} in which Meissner et al.\ proposed a model independent method to extract the resonance parameters on the lattice directly from the imaginary-time correlation function of the field operators at finite imaginary-times.
Their idea is similar to ours in the sense that they proposed to determine resonance parameters by fitting the imaginary-time correlation function by  its parametrized form in the small $\mu$ region, in our terminology.
However, our method is superior than theirs in the respect that we parameterize the correlation solely in terms of pole terms as a function of the uniformization variable based on Mittag-Leffler theorem whereas they parameterize the correlation function by the complex energy of an isolated pole and coefficients of the Taylor expansion at the threshold as a function of the energy and also that our method can be applied to not only single-channel but also couple-channel problems by employing uniformization variable for coupled-channels whereas their method can be applied only to a single-channel problem as far as we can see.   

Wren Yamada is grateful for the inspiring research environment fostered at iTHEMS.
Koichi Yazaki would like to thank RIKEN iTHEMS and its former director, Tetsuo Hatsuda,
and present director, Satoshi Iso, for allowing him to be its member and the wonderful atmosphere for research.
The work of Osamu Morimatsu and Toru Sato was supported by JSPS KAKENHI Grant Number JP24K07062.
\bibliography{paper}

@article{CP-PACS:2007wro,
    author = "Aoki, S. and others",
    collaboration = "CP-PACS",
    title = "{Lattice QCD Calculation of the rho Meson Decay Width}",
    eprint = "0708.3705",
    archivePrefix = "arXiv",
    primaryClass = "hep-lat",
    doi = "10.1103/PhysRevD.76.094506",
    journal = "Phys. Rev. D",
    volume = "76",
    pages = "094506",
    year = "2007"
}

@article{Luscher:1985dn,
    author = "L{\"{u}}scher, M.",
    title = "{Volume Dependence of the Energy Spectrum in Massive Quantum Field Theories. 1. Stable Particle States}",
    reportNumber = "DESY-85-144",
    doi = "10.1007/BF01211589",
    journal = "Commun. Math. Phys.",
    volume = "104",
    pages = "177",
    year = "1986"
}

@article{Luscher:1986pf,
    author = "L{\"{u}}scher, M.",
    title = "{Volume Dependence of the Energy Spectrum in Massive Quantum Field Theories. 2. Scattering States}",
    reportNumber = "DESY-86-034",
    doi = "10.1007/BF01211097",
    journal = "Commun. Math. Phys.",
    volume = "105",
    pages = "153--188",
    year = "1986"
}

@article{Luscher:1990ux,
    author = "L{\"{u}}scher, Martin",
    title = "{Two particle states on a torus and their relation to the scattering matrix}",
    reportNumber = "DESY-90-131",
    doi = "10.1016/0550-3213(91)90366-6",
    journal = "Nucl. Phys. B",
    volume = "354",
    pages = "531--578",
    year = "1991"
}

@article{Luscher:1991cf,
    author = "L{\"{u}}scher, Martin",
    title = "{Signatures of unstable particles in finite volume}",
    reportNumber = "DESY-91-052",
    doi = "10.1016/0550-3213(91)90584-K",
    journal = "Nucl. Phys. B",
    volume = "364",
    pages = "237--251",
    year = "1991"
}

@article{Briceno:2017max,
    author = "Briceno, Raul A. and Dudek, Jozef J. and Young, Ross D.",
    title = "{Scattering processes and resonances from lattice QCD}",
    eprint = "1706.06223",
    archivePrefix = "arXiv",
    primaryClass = "hep-lat",
    reportNumber = "JLAB-THY-17-2495, ADP-17-28-T1034",
    doi = "10.1103/RevModPhys.90.025001",
    journal = "Rev. Mod. Phys.",
    volume = "90",
    number = "2",
    pages = "025001",
    year = "2018"
}

@article{BaryonScatteringBaSc:2023zvt,
    author = "Bulava, John and others",
    collaboration = "Baryon Scattering (BaSc)",
    title = "{Two-Pole Nature of the \ensuremath{\Lambda}(1405) resonance from Lattice QCD}",
    eprint = "2307.10413",
    archivePrefix = "arXiv",
    primaryClass = "hep-lat",
    reportNumber = "MIT-CTP/5579",
    doi = "10.1103/PhysRevLett.132.051901",
    journal = "Phys. Rev. Lett.",
    volume = "132",
    number = "5",
    pages = "051901",
    year = "2024"
}

@article{BaryonScatteringBaSc:2023ori,
    author = "Bulava, John and others",
    collaboration = "Baryon Scattering (BaSc)",
    title = "{Lattice QCD study of \ensuremath{\pi}\ensuremath{\Sigma}-K\textasciimacron{}N scattering and the \ensuremath{\Lambda}(1405) resonance}",
    eprint = "2307.13471",
    archivePrefix = "arXiv",
    primaryClass = "hep-lat",
    reportNumber = "MIT-CTP/5580",
    doi = "10.1103/PhysRevD.109.014511",
    journal = "Phys. Rev. D",
    volume = "109",
    number = "1",
    pages = "014511",
    year = "2024"
}

@article{Yamada:2020rpd,
    author = "Yamada, W. and Morimatsu, O.",
    title = "{New method to extract information of near-threshold resonances: Uniformized Mittag-Leffler expansion of Green's function and $T$ matrix}",
    eprint = "2005.07022",
    archivePrefix = "arXiv",
    primaryClass = "hep-ph",
    doi = "10.1103/PhysRevC.102.055201",
    journal = "Phys. Rev. C",
    volume = "102",
    number = "5",
    pages = "055201",
    year = "2020"
}

@article{Yamada:2021cjo,
    author = "Yamada, Wren A. and Morimatsu, Osamu",
    title = "{Application of the uniformized Mittag-Leffler expansion to $\Lambda(1405)$}",
    eprint = "2101.10029",
    archivePrefix = "arXiv",
    primaryClass = "hep-ph",
    reportNumber = "KEK-TH-2291, J-PARC-TH-0237",
    doi = "10.1103/PhysRevC.103.045201",
    journal = "Phys. Rev. C",
    volume = "103",
    number = "4",
    pages = "045201",
    year = "2021"
}

@article{Yamada:2022xam,
    author = "Yamada, Wren A. and Morimatsu, Osamu and Sato, Toru",
    title = "{Analytic Map of Three-Channel S Matrix: Generalized Uniformization and Mittag-Leffler Expansion}",
    eprint = "2203.17069",
    archivePrefix = "arXiv",
    primaryClass = "hep-ph",
    reportNumber = "KEK-TH-2408, J-PARC-TH-0271",
    doi = "10.1103/PhysRevLett.129.192001",
    journal = "Phys. Rev. Lett.",
    volume = "129",
    number = "19",
    pages = "192001",
    year = "2022"
}

@article{Yamada:2021azg,
    author = "Yamada, Wren A. and Morimatsu, Osamu and Sato, Toru and Yazaki, Koichi",
    title = "{Near-threshold spectrum from a uniformized Mittag-Leffler expansion: Pole structure of the Z(3900)}",
    eprint = "2108.11605",
    archivePrefix = "arXiv",
    primaryClass = "hep-ph",
    reportNumber = "KEK-TH-2344, J-PARC-TH-0248, RIKEN-iTHEMS-Report-21",
    doi = "10.1103/PhysRevD.105.014034",
    journal = "Phys. Rev. D",
    volume = "105",
    number = "1",
    pages = "014034",
    year = "2022"
}

@article{Yamada:2023wqp,
    author = "Yamada, Wren A. and Morimatsu, Osamu and Sato, Toru and Yazaki, Koichi",
    title = "{Survival probability of unstable states in coupled-channels: Nonexponential decay of threshold cusp}",
    eprint = "2305.11695",
    archivePrefix = "arXiv",
    primaryClass = "hep-ph",
    reportNumber = "KEK-TH-2526, J-PARC-TH-0287, RIKEN-iTHEMS-Report-23",
    doi = "10.1103/PhysRevD.108.L071502",
    journal = "Phys. Rev. D",
    volume = "108",
    number = "7",
    pages = "L071502",
    year = "2023"
}

@article{Yamada:2025lkc,
    author = "Yamada, Wren and Morimatsu, Osamu and Sato, Toru and Yazaki, Koichi",
    title = "{Pole-expansion of two-hadron imaginary-time correlation function: A new analysis method for unstable states in lattice QCD}",
    eprint = "2505.02878",
    archivePrefix = "arXiv",
    primaryClass = "hep-lat",
    reportNumber = "KEK-TH-2717, J-PARC-TH-0316, RIKEN-iTHEMS-Report-25",
    doi = "10.1103/yjz9-9d8y",
    journal = "Phys. Rev. D",
    volume = "112",
    number = "3",
    pages = "034040",
    year = "2025"
}

@book{Newton:book,
	author = {Roger G. Newton},
	doi = {10.1007/978-3-642-88128-2},
	publisher = {Springer},
	subtitle = {Second Edition},
	title = {Scattering Theory of Waves and Particles},
	year = {1982}}

@article{NEWTON195829,
title = "Inelastic scattering",
journal = "Annals of Physics",
volume = "4",
number = "1",
pages = "29 - 56",
year = "1958",
issn = "0003-4916",
doi = "https://doi.org/10.1016/0003-4916(58)90036-8",
url = "http://www.sciencedirect.com/science/article/pii/0003491658900368",
author = "Roger G Newton"
}

@article{Kato:1965iee,
    author = "Kato, Masaaki",
    title = "{Analytical properties of two-channel S-matrix}",
    doi = "10.1016/0003-4916(65)90235-6",
    journal = "Annals Phys.",
    volume = "31",
    number = "1",
    pages = "130--147",
    year = "1965"
}

@book{arfken2013mathematical,
	author = {Arfken, G.B. and Weber, H.J. and Harris, F.E.},
	isbn = {9780123846549},
	lccn = {2012418308},
	publisher = {Elsevier Science},
	title = {Mathematical Methods for Physicists: A Comprehensive Guide},
	url = {https://books.google.co.jp/books?id=qLFo\_Z-PoGIC},
	year = {2013}}

@book{whittaker1928course,
  author = {Whittaker, Edmund Taylor and Watson, George Neville},
  publisher = {University Press},
  title={A Course of Modern Analysis: an Introduction to the General Theory of Infinite Processes and of Analytic Functions; with an Account of the Principal Transcendental Functions},
  author={Whittaker, ET and Watson, GN},
  year={1928}
}

@article{Dudek:2014qha,
    author = "Dudek, Jozef J. and Edwards, Robert G. and Thomas, Christopher E. and Wilson, David J.",
    collaboration = "Hadron Spectrum",
    title = "{Resonances in coupled $\pi K -\eta K$ scattering from quantum chromodynamics}",
    eprint = "1406.4158",
    archivePrefix = "arXiv",
    primaryClass = "hep-ph",
    reportNumber = "JLAB-THY-14-1891, DAMTP-2014-36",
    doi = "10.1103/PhysRevLett.113.182001",
    journal = "Phys. Rev. Lett.",
    volume = "113",
    number = "18",
    pages = "182001",
    year = "2014"
}

@article{Wilson:2015dqa,
    author = "Wilson, David J. and Briceno, Raul A. and Dudek, Jozef J. and Edwards, Robert G. and Thomas, Christopher E.",
    title = "{Coupled $\pi\pi, K\bar{K}$ scattering in $P$-wave and the $\rho$ resonance from lattice QCD}",
    eprint = "1507.02599",
    archivePrefix = "arXiv",
    primaryClass = "hep-ph",
    reportNumber = "DAMTP-2015-34, JLAB-THY-15-2101",
    doi = "10.1103/PhysRevD.92.094502",
    journal = "Phys. Rev. D",
    volume = "92",
    number = "9",
    pages = "094502",
    year = "2015"
}

@article{Dudek:2016cru,
    author = "Dudek, Jozef J. and Edwards, Robert G. and Wilson, David J.",
    collaboration = "Hadron Spectrum",
    title = "{An $a_0$ resonance in strongly coupled $\pi \eta$, $K\overline{K}$ scattering from lattice QCD}",
    eprint = "1602.05122",
    archivePrefix = "arXiv",
    primaryClass = "hep-ph",
    reportNumber = "JLAB-THY-16-2216, DAMTP-2016-18",
    doi = "10.1103/PhysRevD.93.094506",
    journal = "Phys. Rev. D",
    volume = "93",
    number = "9",
    pages = "094506",
    year = "2016"
}

@article{Moir:2016srx,
    author = "Moir, Graham and Peardon, Michael and Ryan, Sin\'ead M. and Thomas, Christopher E. and Wilson, David J.",
    title = "{Coupled-Channel $D\pi$, $D\eta$ and $D_{s}\bar{K}$ Scattering from Lattice QCD}",
    eprint = "1607.07093",
    archivePrefix = "arXiv",
    primaryClass = "hep-lat",
    reportNumber = "DAMTP-2016-48",
    doi = "10.1007/JHEP10(2016)011",
    journal = "JHEP",
    volume = "10",
    pages = "011",
    year = "2016"
}

@article{Alexandrou:2017mpi,
    author = "Alexandrou, Constantia and Leskovec, Luka and Meinel, Stefan and Negele, John and Paul, Srijit and Petschlies, Marcus and Pochinsky, Andrew and Rendon, Gumaro and Syritsyn, Sergey",
    title = "{$P$-wave $\pi\pi$ scattering and the $\rho$ resonance from lattice QCD}",
    eprint = "1704.05439",
    archivePrefix = "arXiv",
    primaryClass = "hep-lat",
    reportNumber = "RBRC-1234",
    doi = "10.1103/PhysRevD.96.034525",
    journal = "Phys. Rev. D",
    volume = "96",
    number = "3",
    pages = "034525",
    year = "2017"
}

@article{Andersen:2018mau,
    author = {Andersen, Christian and Bulava, John and H\"orz, Ben and Morningstar, Colin},
    title = "{The $I=1$ pion-pion scattering amplitude and timelike pion form factor from $N_{\rm f} = 2+1$ lattice QCD}",
    eprint = "1808.05007",
    archivePrefix = "arXiv",
    primaryClass = "hep-lat",
    reportNumber = "CP3-Origins-2018-029 DNRF90 MITP/18-073",
    doi = "10.1016/j.nuclphysb.2018.12.018",
    journal = "Nucl. Phys. B",
    volume = "939",
    pages = "145--173",
    year = "2019"
}

@article{CS:2011vqf,
    author = "Aoki, S. and others",
    collaboration = "CS",
    title = "{$\rho$ Meson Decay in 2+1 Flavor Lattice QCD}",
    eprint = "1106.5365",
    archivePrefix = "arXiv",
    primaryClass = "hep-lat",
    doi = "10.1103/PhysRevD.84.094505",
    journal = "Phys. Rev. D",
    volume = "84",
    pages = "094505",
    year = "2011"
}

@article{Dudek:2012xn,
    author = "Dudek, Jozef J. and Edwards, Robert G. and Thomas, Christopher E.",
    collaboration = "Hadron Spectrum",
    title = "{Energy dependence of the $\rho$ resonance in $\pi\pi$ elastic scattering from lattice QCD}",
    eprint = "1212.0830",
    archivePrefix = "arXiv",
    primaryClass = "hep-ph",
    reportNumber = "JLAB-THY-12-1666, TCDMATH-12-10",
    doi = "10.1103/PhysRevD.87.034505",
    journal = "Phys. Rev. D",
    volume = "87",
    number = "3",
    pages = "034505",
    year = "2013",
    note = "[Erratum: Phys.Rev.D 90, 099902 (2014)]"
}

@article{Feng:2010es,
    author = "Feng, Xu and Jansen, Karl and Renner, Dru B.",
    title = "{Resonance Parameters of the rho-Meson from Lattice QCD}",
    eprint = "1011.5288",
    archivePrefix = "arXiv",
    primaryClass = "hep-lat",
    reportNumber = "DESY-10-176, SFB-CPP-10-117, MS-TP-10-14, KEK-CP-242, JLAB-THY-10-1289",
    doi = "10.1103/PhysRevD.83.094505",
    journal = "Phys. Rev. D",
    volume = "83",
    pages = "094505",
    year = "2011"
}

@article{Lang:2011mn,
    author = "Lang, C. B. and Mohler, Daniel and Prelovsek, Sasa and Vidmar, Matija",
    title = "{Coupled channel analysis of the rho meson decay in lattice QCD}",
    eprint = "1105.5636",
    archivePrefix = "arXiv",
    primaryClass = "hep-lat",
    doi = "10.1103/PhysRevD.89.059903",
    journal = "Phys. Rev. D",
    volume = "84",
    number = "5",
    pages = "054503",
    year = "2011",
    note = "[Erratum: Phys.Rev.D 89, 059903 (2014)]"
}

@article{Chen:2022asf,
    author = "Chen, Hua-Xing and Chen, Wei and Liu, Xiang and Liu, Yan-Rui and Zhu, Shi-Lin",
    title = "{An updated review of the new hadron states}",
    eprint = "2204.02649",
    archivePrefix = "arXiv",
    primaryClass = "hep-ph",
    doi = "10.1088/1361-6633/aca3b6",
    journal = "Rept. Prog. Phys.",
    volume = "86",
    number = "2",
    pages = "026201",
    year = "2023"
}

@article{Mai:2022eur,
    author = "Mai, Maxim and Mei\ss{}ner, Ulf-G. and Urbach, Carsten",
    title = "{Towards a theory of hadron resonances}",
    eprint = "2206.01477",
    archivePrefix = "arXiv",
    primaryClass = "hep-ph",
    doi = "10.1016/j.physrep.2022.11.005",
    journal = "Phys. Rept.",
    volume = "1001",
    pages = "1--66",
    year = "2023"
}

@article{DeWitt:1956be,
    author = "DeWitt, Bryce S.",
    title = "{Transition from discrete to continuous spectra}",
    doi = "10.1103/PhysRev.103.1565",
    journal = "Phys. Rev.",
    volume = "103",
    pages = "1565--1571",
    year = "1956"
}

@article{Meissner:2010rq,
    author = "Meissner, Ulf-G. and Polejaeva, Kathryn and Rusetsky, Akaki",
    title = "{Extraction of the resonance parameters at finite times}",
    eprint = "1007.0860",
    archivePrefix = "arXiv",
    primaryClass = "hep-lat",
    reportNumber = "HISKP-TH-10-16, FZJ-IKP(TH)-2010-12",
    doi = "10.1016/j.nuclphysb.2010.12.011",
    journal = "Nucl. Phys. B",
    volume = "846",
    pages = "1--20",
    year = "2011"
}

@article{Doring:2011vk,
    author = "Doring, M. and Meissner, Ulf-G. and Oset, E. and Rusetsky, A.",
    title = "{Unitarized Chiral Perturbation Theory in a finite volume: Scalar meson sector}",
    eprint = "1107.3988",
    archivePrefix = "arXiv",
    primaryClass = "hep-lat",
    doi = "10.1140/epja/i2011-11139-7",
    journal = "Eur. Phys. J. A",
    volume = "47",
    pages = "139",
    year = "2011"
}

@article{Kaplan:1996xu,
	archiveprefix = {arXiv},
	author = {Kaplan, David B. and Savage, Martin J. and Wise, Mark B.},
	doi = {10.1016/0550-3213(96)00357-4},
	eprint = {nucl-th/9605002},
	journal = {Nucl. Phys. B},
	pages = {629--659},
	reportnumber = {DOE-ER-40561-257, INT-96-00-125, UW-PT-96-06, CMU-HEP-96-06, DOE-ER-40862-117, CALT-68-2047},
	title = {{Nucleon - nucleon scattering from effective field theory}},
	volume = {478},
	year = {1996}
}

@article{Yamaguchi:1954mp,
    author = "Yamaguchi, Yoshio",
    title = "{Two nucleon problem when the potential is nonlocal but separable. 1.}",
    doi = "10.1103/PhysRev.95.1628",
    journal = "Phys. Rev.",
    volume = "95",
    pages = "1628--1634",
    year = "1954"
}
\end{document}